\documentclass[traditabstract]{aa}

\usepackage{epsfig}
\usepackage{graphics}
\usepackage[latin1]{inputenc}
\usepackage{natbib}

\begin{document}
\title {Photometric selection of Type Ia supernovae 
in the Supernova Legacy Survey}
\author{G.~Bazin\inst{1,17}, V.~Ruhlmann-Kleider\inst{1}, 
N.~Palanque-Delabrouille\inst{1}, J.~Rich\inst{1}, 
E.~Aubourg\inst{1},
P.~Astier\inst{2}, C.~Balland\inst{2,3},
S.~Basa\inst{4},
R.~G.~Carlberg\inst{5}, A.~Conley\inst{6}, D.~Fouchez\inst{7}, J.~Guy\inst{2},
D.~Hardin\inst{2}, I.~M.~Hook\inst{8,9}, D.~A.~Howell\inst{10,11},
R.~Pain\inst{2}, K.~Perrett\inst{5,12},
C.~J.~Pritchet\inst{13}, N.~Regnault\inst{2},
M.~Sullivan\inst{8}, N.~Fourmanoit\inst{2}, 
S.~Gonz\'alez-Gait\'an\inst{5}, C.~Lidman\inst{14}, 
S.~Perlmutter\inst{15},
P.~Ripoche\inst{15,2}, 
E.~S.~Walker\inst{8,16}
}
\institute{CEA, Centre de Saclay, Irfu/SPP,  91191 Gif-sur-Yvette, France
\and  LPNHE, Universit\'e Pierre et Marie Curie, Universit\'e Paris Diderot, 
CNRS-IN2P3, 4 place Jussieu, 75252 Paris Cedex 05, France
\and University Paris 11, Orsay, 91405, France
\and LAM, CNRS-INSU, BP8, P\^{o}le de l'\'etoile, Site de Ch\^{a}teau-Gombert,
38 rue Fr\'ed\'eric Joliot-Curie, 13388 Marseille Cedex 13, France
\and Deparment of Astronomy and Astrophysics, University of Toronto, 
50 St. George Street, Toronto, ON M5S 3H8, Canada
\and Center for Astrophysics and Space Astronomy, University of Colorado, 
Boulder, CO 80309-0389, USA
\and CPPM, CNRS-Luminy, Case 907, 13288 Marseille Cedex 9, France
\and Department of Physics (Astrophysics), University of Oxford, 
Denys Wilkinson Building, Keble Road, Oxford OX1 3RH, UK
\and INAF - Osservatorio Astronomico di Roma, via Frascati 33, 
00040 Monteporzio (RM), Italy
\and Las Cumbres Observatory Global Telescope Network, 6740 Cortona Dr.,
Suite 102, Goleta, CA 93117, USA
\and Department of Physics, University of California, Santa Barbara, 
Broida Hall, Mail Code 9530, Santa Barbara, CA 93106-9530, USA
\and Network Information Operations, DRDC-Ottawa, 3701 Carling Avenue, 
Ottawa, ON, K1A 0Z4, Canada
\and Department of Physics and Astronomy, University of Victoria, 
PO Box 3055, Victoria, BC V8W 3P6, Canada
\and Australian Astronomical Observatory, P.O. Box 296, Epping, NSW 1710, Australia
\and LBNL, 1 Cyclotron Rd, Berkeley, CA 94720, USA
\and Scuola Normale Superiore, Piazza des Cavalieri 7, 56126 Pisa, Italy
\and Department of Physics, Ludwig-Maximilians-Universit\"at, 
Scheinerstr. 1, 81679 M\"unchen
and Excellence Cluster Universe, Boltzmannstr. 2, 85748 Garchhing, Germany
}
\date{Received 16 March 2011; accepted 23 August 2011}
\authorrunning{G. Bazin et al.}
\titlerunning{Photometric selection of Type Ia supernovae 
in the SNLS}
\abstract{
We present a sample of 485 photometrically identified Type Ia supernova
candidates mined from the first three years of data of the CFHT 
SuperNova Legacy Survey (SNLS). The images were submitted to 
a deferred processing independent of the SNLS real-time detection pipeline. 
Light curves of all transient events were reconstructed  in the $g_M$, $r_M$, 
$i_M$ and $z_M$ filters and submitted to automated sequential cuts in order 
to identify possible supernovae.
Pure noise and long-term variable events were rejected by light curve
shape criteria.
Type Ia supernova identification relied on event characteristics fitted to 
their light curves assuming the events to be normal SNe~Ia. 
The light curve fitter SALT2 was used for this purpose, assigning host galaxy 
photometric redshifts to the tested events.
The selected sample of 485 candidates is one magnitude deeper than 
that allowed by the SNLS spectroscopic identification. 
The contamination by supernovae of other types is estimated to be 4\%.
Testing Hubble diagram residuals with this enlarged sample allows us
to measure the Malmquist bias due to spectroscopic selections directly. 
The result is fully consistent with the precise Monte Carlo based estimate 
used to correct SN~Ia distance moduli in the SNLS 3-year cosmological analyses.
This paper demonstrates the feasibility of a photometric selection of 
high redshift supernovae with known host galaxy redshifts, opening 
interesting prospects for cosmological analyses from future large photometric 
SN~Ia surveys.
}
\maketitle

\section{Introduction}
\par The accelerated expansion of the Universe was revealed
by studies of high redshift Type~Ia supernovae 
(SNe Ia)~\citep{bib:riess, bib:perlmutter}.
Since then, second-generation experiments with efficient observation strategies
were set up to achieve more precise cosmological measurements from SNe~Ia, 
such as the SNLS~\citep{bib:astier},
ESSENCE~\citep{bib:essence} and 
SDSS-II Supernova Survey projects~\citep{bib:sdss}. 
In order to meet the precision required by such measurements, 
samples of SNe~Ia rely on photometry to detect transient events and to measure
their light curves, while  SN~Ia types and redshifts are provided by follow-up
spectroscopy.
In this paper, we describe a method based on photometry alone to detect and 
select supernova candidates, assigning them galaxy photometric redshifts from an 
external catalogue. This method was applied in a deferred analysis of the first 
three years of the Supernova Legacy Survey (SNLS) conducted at the Canada-France-Hawaii 
Telescope (CFHT) from 2003 to 2008, as part of the Deep Synoptic Survey 
of the CFHT Legacy Survey (CFHT-LS).

\par 
SNLS was defined with an optimised observing strategy.
Four one-square degree fields were targeted throughout the 5 
to 7 consecutive lunations where they remained visible.
In every lunation, each visible field was repeatedly observed every 3 
or 4 nights during dark time. 
This rolling-search mode allowed supernova light curves 
to be measured with very good time sampling in four broadband filters,
denoted $g_M$, $r_M$, $i_M$ and $z_M$, similar to the SDSS filter 
set~\citep{bib:regnault} and spanning the wavelength range from 400 to 1000~nm. 
These measurements were made with the MegaCam imager~\citep{bib:boulade}, 
a 1 square degree array of 36 CCDs with 340 million pixels in total.

The standard SN~Ia selection for the SNLS cosmological analyses, 
based on real-time detection~\citep{bib:perrett} and follow-up spectroscopy
can be found 
in~\cite{bib:astier} for the first year data sample and 
in~\cite{bib:guy2010} for the 3-year sample. 
Spectral data from the SNLS follow-up spectroscopy  
are described in~\cite{bib:howell},~\cite{bib:bronder}, \cite{bib:ellis},
\cite{bib:balland} and~\cite{bib:walker}.
Photometric selections are commonly used in SNLS to define spectroscopic
follow-up prioritisation from real-time (and thus partial) light curves~\citep{bib:sullivan}
or to enlarge the real-time SN~Ia sample in order to measure SN~Ia rates or 
properties~\citep{bib:sullivanb}. 
Photometric SN classification is also expected to become a key challenge for
future supernova projects by lack of spectroscopic resources to confirm the  
large amount of supernovae that will be detected~\citep{bib:snchallenge}.

The photometric analysis described in this paper is independent from the SNLS 
real-time detection pipeline and the ensuing cosmological analyses. 
It uses different data processing and detection, and applies a coherent set of cuts 
on 3-year light curves of all detected events, whether with or without spectroscopic 
information. 
Benefitting from more complete light curves and less stringent minimal brightness
constraints than the real-time spectroscopic selection, the photometric analysis
aims at going deeper in magnitude. 
As such, it can provide a cross-check to estimate possible biases of the standard SNLS 
SN~Ia selection for cosmological analyses.
It is also the basis of a photometric selection of non-type Ia supernovae that allowed
the core-collapse supernova rate to be measured at a mean
 redshift $z\sim0.3$~\citep{bib:ratecc}. 
Ultimately, it aims at exploring the feasibility of a cosmological analysis with
a photometric SN~Ia sample, which will be the subject of a future paper.

The outline of the paper is as follows. Data processing, transient
event detection and photometry are described in Section~\ref{sec:detphotsec}.  
The selection of SN-like transient events and their association with host
galaxy photometric redshifts are presented in Section~\ref{sec:presel}.
The selection of SN~Ia candidates is discussed in Section \ref{sec:sn}.  
The characteristics of the SN~Ia candidates
are presented in Section \ref{sec:char} where
selected events with and without spectroscopic identification are compared
in the same range of magnitude and found to have similar properties.
Finally, distance estimates from the whole photometric sample and from
the subsample with spectroscopic identification, which has a lower limiting magnitude,
are compared in Section~\ref{sec:bias}.

\section{Detection and photometry}\label{sec:detphotsec}

\subsection{Transient event detection}\label{sec:transient}
\par Data corresponding to SNLS observations
from March $1^{\rm st}$, 2003 to September $21^{\rm th}$, 2006 
were analysed, using images pre-processed at the end of each 
lunation with the Elixir pipeline at CFHT~\citep{bib:elixir}
which provided flat-fielding and fringe subtraction.
These data include the presurvey period (up to June $1^{\rm st}$, 2003) 
which was used for the commissioning of MegaCam and thus corresponds 
to unstable observing conditions.

\par The astrometric solution for the images of each of the four SNLS
fields (D1, D2 , D3 and D4) was computed with the 
TERAPIX~\footnote{see http://www.astromatic.net/software
for the TERAPIX software} calibration
tool SCAMP~\citep{bib:scamp} and the USNO-B1.0 catalogue~\citep{bib:usno}.
The images were then resampled with the TERAPIX tool SWarp~\citep{bib:swarp}
according to the previously determined astrometric solution. 
A set of about 20 images with the best photometric quality 
(based on seeing and absorption considerations) was selected for each 
field and each filter ($g_M$, $r_M$, $i_M$ or $z_M$) 
and co-added with SWarp in order to define reference images.
The selected images were all chosen in the first or second season, depending on 
the field. The same reference images were used throughout the processing. 

\par Cosmic rays, defects due to dead pixels or induced by the resampling 
procedure were searched for in the resampled images and in the reference
images. The content of the corresponding pixels was replaced by the estimate
of the local sky background.
An image subtraction package, TRITON~\citep{bib:leguillou}, based on the 
algorithm of~\cite{bib:alard}, was then applied on every resampled image. 
In this procedure,
the convolution kernel and the sky background were determined independently 
on eight identical non-overlapping tiles paving each CCD of the mosaic.
The number of tiles was chosen to optimise the precision of the subtractions,
spatial variations of the kernel and sky background requiring small tiles,
whereas the kernel determination requires a sufficient number of bright 
objects and thus large tiles. 
To select bright objects to be used in the convolution kernel determination
for a given tile, 
the TERAPIX tool SExtractor~\citep{bib:sextractor} was applied to 
both the reference and the current images, with a detection threshold 
at 2$\sigma$ w.r.t. sky background. 
Bright objects present in both selections and neither saturated
nor too close to a tile boundary were kept, which left
about 100 objects per tile for images at the average seeing of 
0.7~arcsec.
The kernel and sky background were then fitted 
in 59x59 pixel regions centred on each object. Adjacent objects whose
signal regions were overlapping by more than 20\% were discarded.
Subtractions were considered as valid only if the integral of the convolution 
kernel was found to be above a filter-dependent threshold. 
The rate of lost subtractions (either failed or not valid) was around 
a few \%. 

\par In each field, subtracted images in the $i_M$ filter were  
stacked for each lunation. On average, 27 images entered each stack.
SExtractor was applied on each stack to construct catalogues of objects
exhibiting a variable flux - whether positive or negative - 
in the $i_M$ filter in any lunation.
At least 4 adjacent pixels with a signal of more than 2.5$\sigma$ 
w.r.t. sky background were required to confirm an object. 
As the same set of reference images was kept throughout the whole processing,
some supernovae may have part of their signal included in the references.
In that case, the events might appear as missing flux on each of the images and 
could have been missed if searching only for a flux excess. 
Considering negative measured fluxes when building the above catalogues allowed us 
to recover those events.

All lunation catalogues were then merged to produce the final detection map.
In this way, any object detected on several lunations gave only one detection, with
a position averaged over all lunation stacks. 

\par 
When applied to the 3-year data sample of SNLS, the above detection pipeline
produced a total of about 300,000 detections, dominated mostly by saturated
signals from bright objects which were not perfectly subtracted. 
The efficiency of the pipeline is discussed in the next section, while 
photometry is described in section~\ref{sec:calib}
and further selections to discriminate between spurious detections and
supernova signals are the subject of section~\ref{sec:presel}.

\subsection{Detection efficiency}\label{sec:detec}
In order to check the performance of the above pipeline, we studied the
detection efficiency with Monte-Carlo generated artificial images produced
for the D1 field in the $i_M$ filter that we used for the detection.
From this study we derived an efficiency model that allowed us to determine 
the efficiency for the other fields, D2, D3 and D4, which had slightly different
mean observing conditions. This efficiency model was then coupled with a simulation
of synthetic light curves (described in Section~\ref{sec:lcsimul}) that we set
up to study our photometric selection criteria.

The D1 Monte-Carlo images were obtained by adding 216,000 simulated
supernovae to real images, thus naturally reproducing the observing conditions 
(sampling frequencies  
and photometric quality) of the experiment~\citep{bib:ripoche}. 
An external catalogue of host galaxies identified from deep image stacks of the 
CFHT-LS Deep Fields ~\citep{bib:ilbert2006} was used to randomly 
choose the position of each simulated supernova, following 
the observed surface brightness of the hosts, and to set the redshift, 
equal to the photometric redshift of the host. 
The redshift was restricted to the range between 0.2 and 1.2.
The supernovae were attributed a random stretch, colour and 
intrinsic dispersion term according to the distributions observed with 
spectroscopically identified SNe~Ia. The time of maximum luminosity was drawn 
uniformly within each roughly six-month long season of observation. The $i_M$-band 
light curve of each artificial supernova was computed using 
SALT2~\citep{bib:guy2007}
and the supernova flux at each observation date as deduced from the light curve 
was added to the corresponding images.
The images were then processed with the same pipeline as the real images, 
setting the backgrounds and convolution kernels to those measured 
on the original images in order to avoid biases in the subtraction procedure 
due to the presence of the numerous simulated supernovae. 
The detection efficiency was then defined as the fraction of simulated 
supernovae recovered in the detection map at the end of the processing.

\begin{figure}[hbtp]
\begin{center}
\epsfig{figure=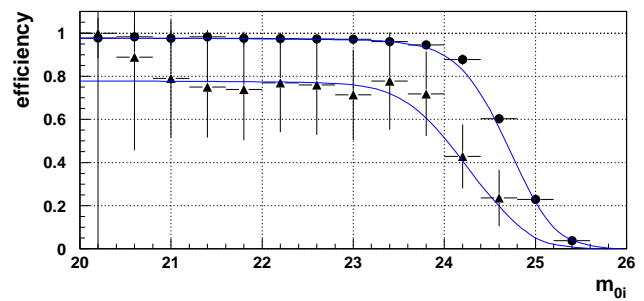,width = \columnwidth} 
\caption[]{Detection efficiency $\epsilon(m_{0i})$ for the first season of D1
as a function of the generated peak magnitude in $i_M$. 
The model (blue curve) is compared with Monte-Carlo estimates (symbols), 
showing separately the efficiency for CCD 3 (triangles), or averaged over the rest 
of the mosaic (dots).} 
\label{fig:effD1_m1}
\end{center}
\end{figure}

The detection efficiency for a supernova with date $t_{0i}$ of maximum light in 
$i_M$-band depends on its peak magnitude $m_{0i}$, on the seeing and sky background 
during nearby observing times $t_k$, and on the relative epochs $t_{0i}-t_k$. 
The Monte Carlo images allowed us to study the efficiency dependence on these variables,
in order to build an efficiency model to extrapolate from D1 to the other fields.
As an example, the average first season detection efficiency, $\epsilon(m_{0i})$
is illustrated in Figure~\ref{fig:effD1_m1} for D1.
The efficiency is rather uniform over the CCD mosaic, except for
CCD 3 which failed to work during three out of the six dark time periods of the 
first season. The efficiency is nearly magnitude-independent,
$\epsilon_{max}\sim 0.97$, out to $m_{0i}=23.5$ in the SNLS magnitude 
system~\footnote{The SNLS magnitude system uses star BD +17 4708 as a flux 
standard, which, as compared with the standard reference Vega, has the advantage
of known Landolt magnitudes, precise SED measurements and colours close to the avergae colours
of the Landolt stars used in the SNLS calibration~\citep{bib:regnault}}.
This is followed  with a steep decline at faint magnitudes reaching an efficiency 
of 0.50 at $m_{0i}\sim24.7$. 
Very similar efficiencies were obtained in the other two seasons. Figure~\ref{fig:effD1_m1} 
also shows the result of our efficiency model that reproduces the dependence of the 
efficiency as a function of the SN~Ia peak magnitude.
This model is described in Appendix~A.

\begin{figure}[htbp]
\begin{center}
\epsfig{figure=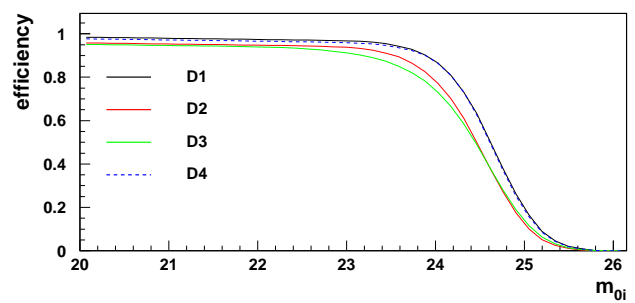,width = \columnwidth} 
\caption[]{Detection efficiency averaged over the entire CCD mosaic for fields 
D1, D2, D3 and D4 as a function of the generated peak magnitude in $i_M$. } 
\label{fig:eff_m_allyr}
\end{center}
\end{figure}

The efficiency model was applied to fields D2, D3 and D4, using the observation 
dates on these fields and the information of seeing and background for each exposure. 
Each CCD was treated independently 
to reproduce exactly the possible failures, whether electronic or due to the 
processing of the real images. The detection efficiency computed for each 
field over the observing period between March 2003 and September 2006 and 
averaged over the entire mosaic is shown in Figure~\ref{fig:eff_m_allyr}. 
Fields D2 and D3 have a smaller efficiency than D1 and D4. This is caused 
by the lack of data due to bad weather in February 2004 and in March 2006, 
when only D2 and D3 were observable.

\subsection{Photometry and calibration}\label{sec:calib}
For each of the 300,000 detections, four-filter light curves 
were built from individual subtracted images.
TRITON differential photometry with Point Spread Function (PSF) fitting
was applied, imposing the position found in the $i_M$ detection map 
(see Section~\ref{sec:transient}).
Fluxes were normalised by the integral of the convolution kernel determined 
in the subtractions, in order to express all fluxes on a common scale, that 
of the reference image. 
For each exposure and CCD, the PSF was determined as follows.
Bright events were first detected on each resampled image using 
SExtractor. The flux in the pixel of maximum content
and the integrated flux of the event allowed saturated stars and galaxies 
to be identified and rejected. This left about 100 non saturated stars 
in each CCD, which were used to define one PSF per CCD by averaging 
the star profiles computed in a seeing-dependent box. 

The measured fluxes were calibrated using the SNLS tertiary standards published
in~\cite{bib:regnault}. 
We measured the tertiary star fluxes directly on current images with the same
photometry algorithm and PSF as for transient events and, as for the latter, the
star fluxes were normalised by the convolution kernel integral determined 
in the subtractions.
Using the magnitudes published in~\cite{bib:regnault}, 
the tertiary star fluxes were then converted into zero-points 
for each exposure and a weighted average of these zero-points was computed over 
each CCD, rejecting zero-points above 5$\sigma$ w.r.t. the median value.
The measured star fluxes were compared with fluxes derived from the star magnitudes by 
applying the average zero-points and the relative flux difference was fitted as a function
of flux (assuming a linear variation). 
In the whole range of tertiary star fluxes, the fitted relative flux difference was found to
remain below 0.3\% whatever the field or band.
Note that~\cite{bib:regnault} also provides flat-fielding corrections due to 
residual radial non-uniformities of the imager photometric response. These
corrections were applied to all measured fluxes, prior to any further treatment 
(conversion into zero-point for stars or calibration for transient events).

The calibrated fluxes were then filtered. 
Residual defects from bad pixels or cosmic rays, detections due to satellite trails or 
saturated stars were identified on subtracted images as bright objects selected
by SExtractor with a detection threshold at 3$\sigma$ w.r.t. sky background. 
Photometric points whose PSF overlapped the spatial extension of any of the
above objects were discarded.
Flux measurements obtained under bad seeing conditions (above 1.2$^{\prime\prime}$)
were also eliminated, as well as flux measurements in nights where all exposures 
were of bad quality~\footnote{grade C from the TERAPIX visual image quality 
control as found in the TERAPIX Spica database, http://clix.iap.fr/steeringgroup/qf/}.
At this point, flux errors were renormalised so that after filtering
the distribution of the pulls of the exposure fluxes w.r.t. the night 
flux average be of variance 1 in the test-sample
of 278 light curves matched to spectroscopically identified SNe~Ia in
our detections (see Section~\ref{sec:lcrta}). Then, in any light curve,
points with flux errors either too low (w.r.t. the expected flux uncertainty 
due to sky background) or too high were discarded. 
Only exposures with a flux within 3$\sigma$ from the night flux median 
value were retained and only nights with at least two such exposures were kept.
On average over all reconstructed light curves, the fraction of measurements removed 
by the above cuts is at most 15\%.

\par Flux measurements corresponding to 
selected exposures within the same night were weighted by their errors 
and averaged to define mean light curves in each of the four filters.
Finally, in order to account for a possible contribution of variable objects
in the reference images, a common baseline, 
defined as the mode of the flux distribution,
was computed in each mean light curve and then withdrawn from the latter.

\begin{table}[htb]
\caption[]{Performance of the photometry used in this work.}
\label{tab:testm}
\begin{center}
\begin{tabular}{ccccc} \hline \hline
Band   & \multicolumn{2}{c}{Full sample} & \multicolumn{2}{c}{Optimal conditions} \\ 
      & Median  & MAD & Median  & MAD \\ \hline
$g_M$ & 0.006 & 0.016 & -0.0001 & 0.014   \\
$r_M$ & 0.018 & 0.028 &  0.006 & 0.020 \\
$i_M$ & 0.015 & 0.023 &  0.001 & 0.016 \\
$z_M$ & 0.009 & 0.054 & -0.002 & 0.047 \\ \hline
\end{tabular}
\tablefoot{The table shows the
difference $m_1-m_2$ of the SALT2 fitted peak magnitudes obtained with the 
photometry of this paper ($m_1$) and with the SNLS-3year photometry 
($m_2$)~:
median magnitude difference and median absolute dispersion (MAD) for 
all SNe Ia with good quality light curves (left) and for the subset detected
in optimal conditions, as described in the text
(right). }
\end{center}
\end{table}

The flux estimation was checked with the set of 278 spectroscopically 
identified SNe Ia previously mentioned.
Peak magnitudes at the date of $B$-band maximum light from our light 
curves were compared with those derived from the SNLS-3year light curves 
published in~\cite{bib:guy2010} 
which correspond to a different image processing and an optimised photometry. 
In order to restrict to light curves of good quality, both sets 
were submitted to sampling and S/N cuts as described in~\cite{bib:guy2010}.
The selected light curves were then fitted with SALT2 in each filter
fixing the date of $B$-band maximum light, $X_1$ and colour at
the values obtained from a multi-band fit to the SNLS-3year light curves
(see Section~\ref{sec:salt}).
Only photometric points present in both sets of light curves were considered in 
the fits.

Differences in the peak magnitudes fitted in each band are summarised
in the left-hand column in Table~\ref{tab:testm}. 
Magnitudes from our light curves are fainter by at most 0.02~mag with
dispersions between 0.02 and 0.05~mag.
Two effects explain these differences. Some of the SNe had SN light
included in our reference images, which had no impact on their detection but
affected their photometry. 
Second, even for SNe with no light in the reference images, the magnitude difference 
was observed to be larger when the SN was located near the top or bottom edges of the
mosaic.
Restricting to SNe with no light in the reference images and located far from 
the top/bottom mosaic borders, closer performance between the two photometry 
techniques are obtained, as shown in the right-hand column in Table~\ref{tab:testm}
(column labelled 'optimal conditions').

The dependence of the magnitude difference with the SN location is likely to be
due to our using the position found in the $i_M$ detection stack as the pivot 
of the PSF photometry. This is not optimal, especially on image edges due to 
distortions induced by alignment which are expected to be more important there.
The average $i_M$  position resolution provided by our detection pipeline was found 
to be 0.41~pixel in the full sample of SNe Ia and 0.32~pixel for SNe in optimal
conditions, to be compared with 0.28~pixel in the photometry 
of~\cite{bib:guy2010} which performs a simultaneous fit of SN~Ia positions 
and fluxes~\footnote{ 
Note that this procedure is ideal when a limited number of objects is to be
reconstructed but would be too time consuming when dealing with all transient
events detected by our photometric pipeline}. 
Position measurement inaccuracy leads to underestimated fluxes. Using appendix B 
of~\cite{bib:guy2010}, we found that resolutions of 0.41, 0.32 and 0.28~pixel 
would lead on average to $i_M$ biases of 0.011, 0.007 and 0.005~mag, respectively.
These numbers reproduce the order of magnitude of the effect reported in 
Table~\ref{tab:testm}.

Summarising, when all SNe are considered, fluxes in this analysis are reconstructed 
with resolutions of a few \% and an uncertainty below 2\% on the absolute flux scale,
which is accurate enough to set up a photometric selection.

\subsection{Catalogue of stars and galaxies}\label{sec:ctlg}
The analysis described in this paper uses a catalogue of
nearly 2 million stars and galaxies present in the four SNLS
fields. It combines two sources. 
The first one, published in~\cite{bib:ilbert2006}, provides positions of 
stars and galaxies, as well as photometric redshifts for the latter.
The second was obtained by applying SExtractor on our reference images 
and contains measurements of the object size. 
The two catalogues were combined into one. 
Events common to both catalogues were classified
as star or galaxy based on their classification in each of the catalogues. 
Ambiguous cases of star vs galaxy classification (e.g. at fainter magnitudes) 
were solved with the help of flux and size measurements from our reference 
images. Altogether, the catalogue consists of 94\% galaxies, 5\% stars
and only 1\% ambiguous objects.

The~\cite{bib:ilbert2006} catalogue also provides photometric redshifts for more
than 520,000 galaxies with an AB magnitude brighter than 25 in $i_M$. 
The photometric redshift method was applied to the CFHT-LS multi-colour data of the 
four 
deep fields of the survey. It was trained with around 3,000 VVDS spectra observed with 
VIMOS in one of the CFHT-LS deep fields, in both visible and near infrared bands.
The method reaches a redshift resolution of 0.037 with 3.7\% of catastrophic redshift 
errors for a sample selected at $i_M \le 24$. All galaxy photometric redshifts 
used in the following are from this catalogue.

\subsection{Light curve fitter}\label{sec:salt}
\par Throughout this paper, we use the SALT2 package
as a SN~Ia light curve fitter. The version 
is that described in ~\cite{bib:guy2010} which was trained on
a larger data sample and has a higher resolution colour variation
law than the original version of ~\cite{bib:guy2007}.
SALT2 models the mean evolution of the spectral energy
distribution sequence of SNe~Ia. The model was trained on Branch-normal
supernova data, both multi-band light curves and spectra, covering 
low and high redshifts up to 0.7. 
The resulting flux model of a SN~Ia at a given redshift
is a function of four parameters.
Those are the $B$-band global flux
normalisation factor, $X_0$, the date of $B$-band maximum light and
two intrinsic rest-frame parameters, a stretch-related parameter, 
$X_1$ and a colour parameter, defined as the $(B-V)$ colour offset
at the date of $B$-band maximum light with respect to the average  
colour in the training sample, $C=(B-V)_{max}-<B-V>$. 

\par In this paper, SALT2 was used either to produce synthetic 
SN~Ia light curves assuming a given cosmology or 
to fit observed light curves under the assumption that they come
from a SN~Ia. In that case, the results are magnitudes, either in
the MegaCam filters or in the rest-frame $B$-band. 
Except in Section~\ref{sec:calib},
these magnitudes are estimated from a global fit to photometric points 
in all filters that are expected to provide a significant contribution 
at the event redshift. 

\section{Supernova selection}\label{sec:presel}
\par First, we apply a set of selection criteria to the light curves
of all candidate detections 
in order to reject spurious objects
and select for SN-like events.
The SN selection criteria, described in the next section, were designed 
to be efficient for all types of supernovae. 
Spectroscopically identified supernovae present in our sample 
(see Section~\ref{sec:lcrta}) 
and synthetic SN~Ia light curves (see Section~\ref{sec:lceff})
were used as qualitative guidelines to define the cuts.  

\subsection{Selection criteria}\label{sec:preselcuts}
The selection proceeds in four steps. The light curves were first searched for
a significant flux variation in order to select variable objects.
We then checked that the main variation in
each curve had a shape consistent with that expected from a SN event and
that there was no other flux variation away from the SN-like variation.
Events likely due to stars were then removed. Eventually, sampling requirements 
were applied to ensure good quality light curves were selected. 
These four steps are detailed below.

\subsubsection{Search for flux variations}
In each light curve and filter, a search for significant flux variations was applied.
Variations were due to start with a photometric point of positive flux and 
significance above 1$\sigma$. 
They ended if a point of negative flux had a significance above 1$\sigma$,
if two successive points of positive fluxes had significances below 1$\sigma$
or at season ends, since supernovae are expected to show only short-term variations
lasting typically over three consecutive lunations.
To reduce the amount of spurious detections, the most significant 
variations found in the $i_M$ and $r_M$ light curves 
were required to contain each at least three points and 
to have their dates of maximum flux within 50 days from one another. 
This reduced the number of detections by about a factor 6.

\subsubsection{SN-like variation}
Most events at this stage were either spurious detections or objects 
varying on a longer term than supernovae.
To be considered further, light curves were then required to have a shape
compatible with that of SN-like events. 
To test this, the main variation in each filter was fitted with the 
phenomenological form~:
\begin{equation}
f^k(t)\;=\; A^k \frac{e^{-(t-t_0^k)/\tau_{fall}^k}}
{1+e^{-(t-t_0^k)/\tau_{rise}^k}} \;+ c^k
\label{fitformula}
\end{equation}
where $k$ indicates the filter. Besides a constant, $c^k$,
the fit parameters are $A^k$, which sets the normalisation
of the variable signal,
$\tau_{fall}^k$ and $\tau_{rise}^k$, which define the fall and rise
times of the variation, respectively, 
and $t_0^k$ which is related to the date of maximum by
$t_{max}^k=t_0^k+\tau_{rise}^k \ln (\tau_{fall}^k/\tau_{rise}^k-1)$.
Fits were independent in the four bands and run over the entire
3-year light curves.
While form~\ref{fitformula} has no particular physical motivation, it is
sufficiently general to fit the light curve shape of all types 
of supernovae. 

To reduce the contamination by long-term variable objects
or by random fluctuations, we first required that points away from the 
main variation were compatible with a constant flux. To do so,
points outside the time interval of the main variation in $i_M$ were compared 
with the constant from the previous fit. Their significances w.r.t. that constant 
were added in quadrature over all four filters and the sum was normalised 
to the total number of points in the sum. This defined the following variable,
hereafter referred as to off-variation $\chi^2$~:
\[
\chi^2_{off}\;=\; \frac{1}{\sum_{k}N_k}
\sum_{k} \sum_{j \notin var} \left( \frac{F_{j}^k - c^k}{\sigma_{j}^{k}}\right)^{2}
\]
where $k$ indicates the filter and $F_{j}$ represents the flux as measured at 
point $j$ in the light curve with an error $\sigma_{j}$.
A cut was applied on $\chi^2_{off}$ as a function of the maximum flux observed 
in $i_M$. As a result, about half of the detections resulted in light curves which, 
outside the main variation, were more irregular than those of SNe. 

\par
\begin{figure}[tbp]
\begin{center}
\epsfig{figure=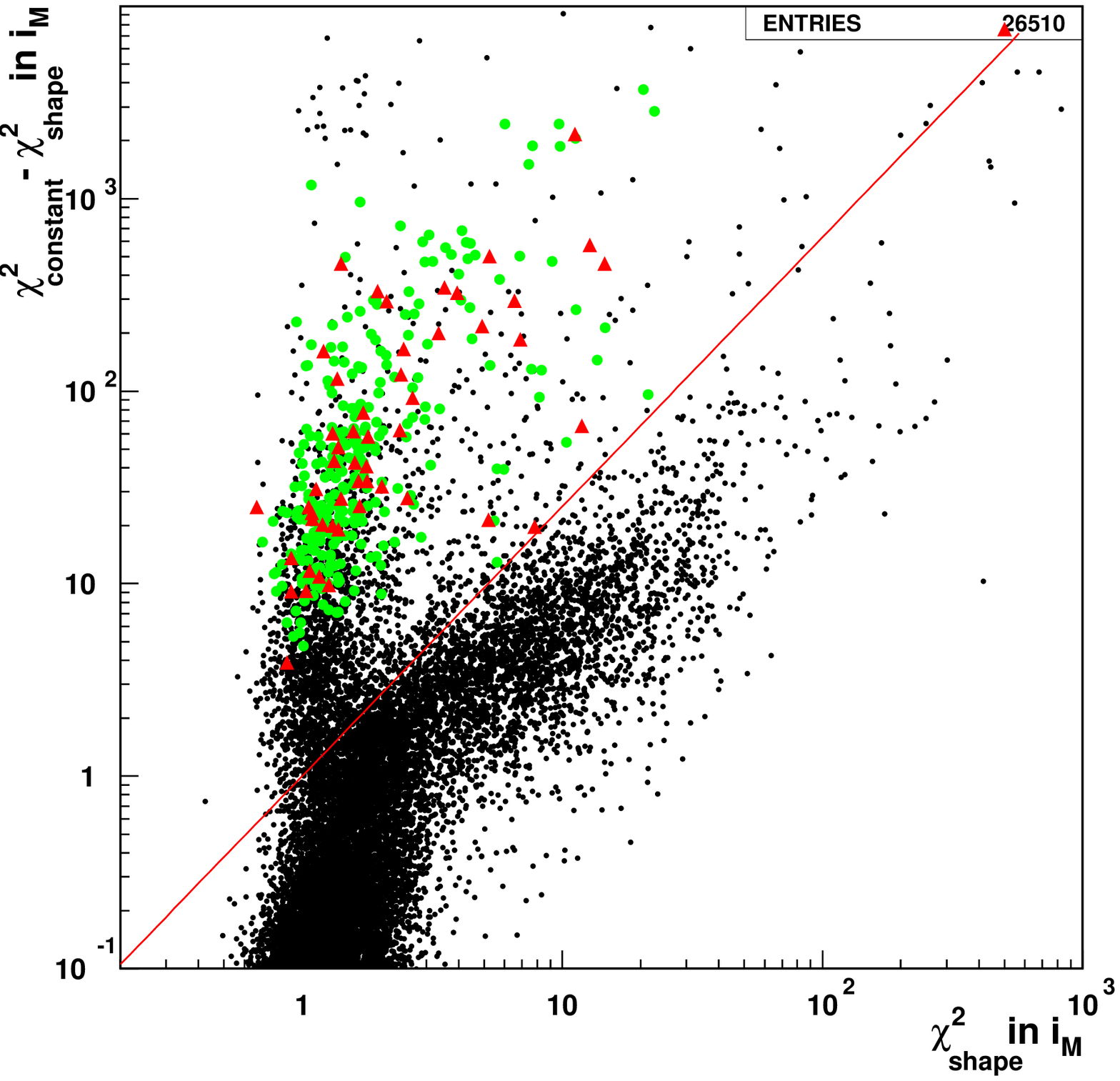,width=\columnwidth}
\epsfig{figure=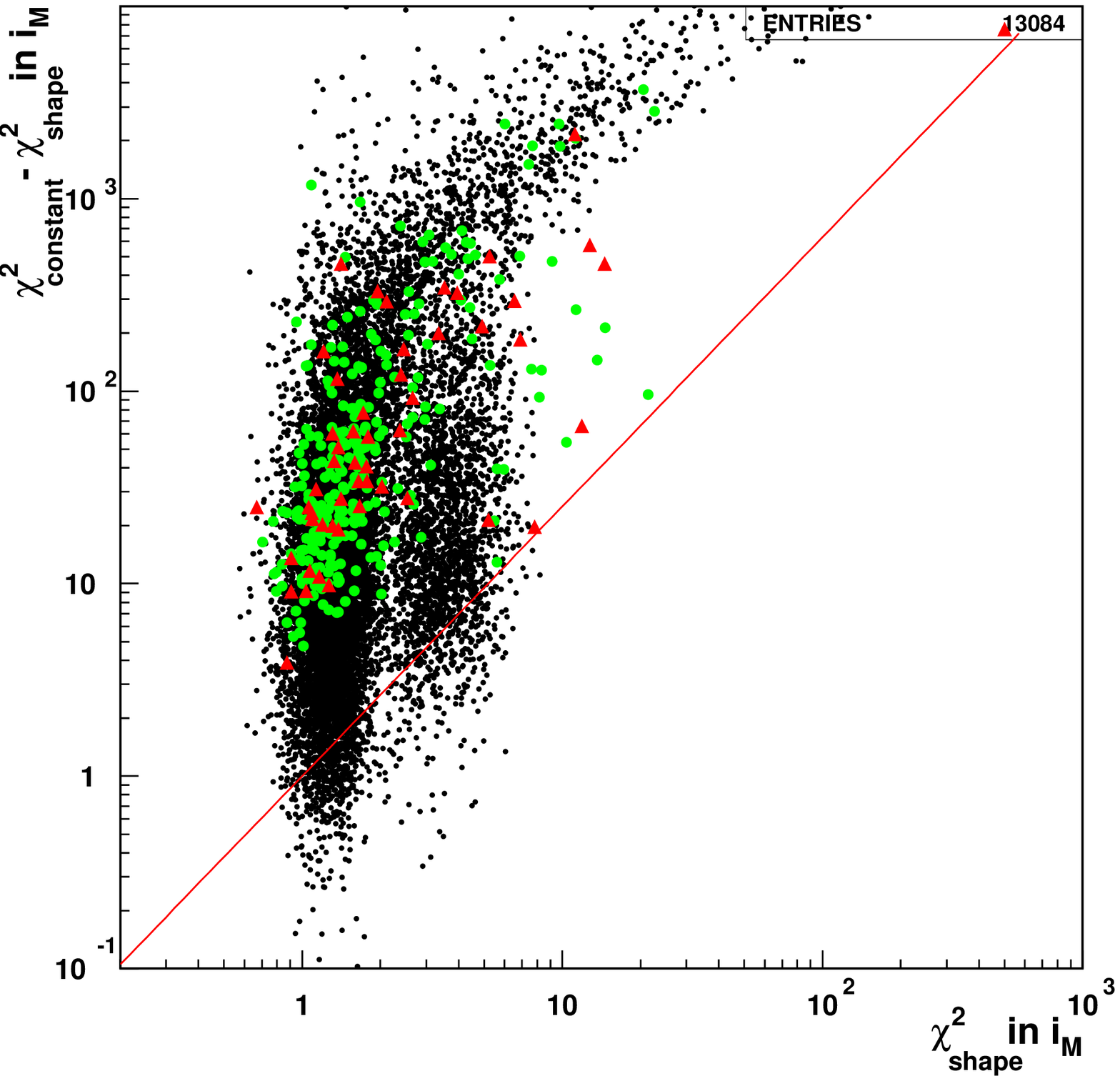, width=\columnwidth}
\caption[]{
Difference in reduced $\chi^2$ between $i_M$ light curve fits 
by a constant and 
by an SN-like shape (see text), as a function of the latter $\chi^2$.
All cuts of the SN selection previous to that illustrated in this
figure have been applied.
Black dots stand for SNLS data (top) and synthetic SN~Ia events (bottom). 
In both plots, green squares (resp. red triangles) are data events
identified by spectroscopy as Type Ia (resp. core-collapse) SNe. Events
above the curve are selected by the analysis. } 
\label{fig:presel2}
\end{center}
\end{figure}

 The shape of the light curve was then tested for consistency 
with (\ref{fitformula}). This was done
in the $i_M$ filter only, as a significant signal is expected in this
filter at all redshifts and for all types of SNe. 
A cut was applied on the difference in reduced $\chi^2$  
between a fit by a constant and the fit by formula~(\ref{fitformula}), 
as a function of the reduced $\chi^2$ of the latter fit. This cut
is illustrated in Figure~\ref{fig:presel2} 
where the events selected by all previous cuts are compared with
simulated SNe~Ia 
(this comparison is further discussed in Section~\ref{sec:simcompa}).
Also highlighted in the plots are the spectroscopically 
identified events contained in our sample.
A clear separation is observed between SN events and the bulk of the 
detections.
After this cut, the number of events was reduced by a factor $\sim$10.
While this cut is more discriminating than that on the off-variation
$\chi^2$, they proved to be complementary, as part ($\sim 12\%$)
of the events rejected by the first cut would pass the second one. 

\subsubsection{Star rejection}
Events at this point were still contaminated by stars, either truly 
variable stars or bright stars leading to saturated signals at some epochs. 
We discarded detections within 3 $\Delta_{PSF}$ from stars in the catalogue 
described in Section~\ref{sec:ctlg}, where $\Delta_{PSF}$ is the 
half-width of the PSF in the $i_M$ reference images.
Saturated signals from bright stars lead to light curves with
large flux changes over a few epochs.
Events whose $i_M$ flux varied by more than 70\% between the date of maximum 
flux and the two closest epochs in the same season were thus rejected.
The above two cuts reduced the number of remaining detections by 12\%.
If they were applied instead of the constraint on the off-variation
$\chi^2$, the reduction factor at that level would be 
similar (46\% instead of 49\%) but the 
sample remaining after the selection on the difference in reduced 
$\chi^2$ would increase by 65\%. 
The three cuts previously described are thus necessary to achieve
a significant rejection against background.

\subsubsection{Sampling requirements}
Quality criteria ensuring sufficient temporal sampling of the
light curves complete the selection. 
In the $i_M$ and $r_M$ bands, at least one pre-max epoch and 
one post-max epoch were required in an interval ranging
from 30 days before the date of maximum flux in each filter
up to 60 days after that date.
These conditions were found appropriate for SN events, even
fast-declining ones like SNe~Ia which on average have time
constants $\tau_{rise}\sim 4$~days and $\tau_{fall}\sim 20$~days
in the $i_M$ and $r_M$ bands.
In the $g_M$ and $z_M$ bands, at least two epochs were
required in a similar interval around the date of maximum 
flux in the $i_M$ filter. 
Finally, the dates of maximum $t_{max}^{i_M}$ and  $t_{max}^{r_M}$
from the fits by formula (\ref{fitformula})
were required to be within 50 days of each other,
as more than 99\% of both the synthetic SNe Ia and the spectroscopically
identifed SN data events were observed to satisfy this condition.

\subsubsection{Results}
At the end of the SN selection, a set of 1483 events is retained. 
The effect of the cuts on data is detailed in Table~\ref{tab:presel},
which also gives the output of the cuts on spectroscopically
identified supernovae and on simulated SN~Ia light curves
(as described in the next two sections).
 
\begin{table}[htb]
\caption[]{Effect of the SN selection cuts.}
\label{tab:presel}
\begin{center}
\begin{tabular}{lrrrr} \hline \hline
Cut & Events & Ia & CC & $\epsilon$(Ia) \\ \hline
Detection in $i_M$      & 295683 & 278 & 55 & 0.96 \\
Variations in $i_M$ 
and $r_M$               & 193104 & 277 & 55 & 0.94 \\
Consistent t$_{max}$ in 
$i_M$ and $r_M$         & 52284  & 276 & 53 & 0.93 \\
Cut in $\chi^2_{off}$   & 26510  & 276 & 52 & 0.93 \\ 
$\chi^2_{\rm constant}-\chi^2_{\rm shape}$ 
cut in $i_M$            &  2474  & 276 & 52 & 0.93 \\ 
Star veto               &  2165  & 275 & 52 & 0.92 \\
Sampling cuts in $i_M, 
r_M$                    &  1674  & 253 & 43 & 0.79 \\
Sampling cuts in $g_M, 
z_M$                    &  1571  & 246 & 42 & 0.73 \\
Consistent t$^{fit}_{max}$ 
in $i_M$ and $r_M$      &  1483  & 246 & 42 & 0.73 \\
\hline
\end{tabular}
\tablefoot{
The table details the effect of the cuts 
on the first three years of SNLS data~:
all detections (column 2), 
subsamples of events identified  by spectroscopy as 
Type Ia and core-collapse SNe (columns 3 and 4). The last
column indicates the efficiency of the cuts for bright 
SNe~Ia, derived from synthetic SN~Ia light curves at low magnitudes
($m_{0i}<23$).} 
\end{center}
\end{table}

\subsection{Comparison with the SNLS real-time selection}\label{sec:lcrta}
The SNLS real-time SN selection procedure is described in~\cite{bib:perrett}. 
It relies on real-time detection of variable events 
based on photometry, combined with a spectroscopic follow-up of the candidates 
which were both likely SNe Ia, as determined from an analysis of their 
real-time light curves~\citep{bib:sullivan},
and bright enough ones to allow for conclusive spectroscopic measurements.
Events fainter than $i_M^{AB} = 24.4$ or far beyond maximum light at 
detection time were not observed spectroscopically. Events with moderate
brightness, $22.9< i_M^{AB} < 24.4$ were qualified for spectroscopy if
their fractional increase in brightness compared with their host galaxy 
was above a magnitude-dependent threshold (see~\cite{bib:perrett} for
more details).
The spectroscopic observations were essential to confirm the nature of the
real-time candidates and to obtain precise redshifts. 
As an example of how efficient the selection of targets for spectroscopy
was, 71\% of the first-year high redshift events sent for 
spectroscopy at the Gemini telescopes were confirmed to be Type Ia supernovae, 
either secure (``SN~Ia'')  or probable Ia events (``SN~Ia?''), as described 
in~\cite{bib:howell}. 

In this section, we compare our photometrically identified SN-like
events to those located by SNLS in real-time.
The SNLS database\footnote{https://legacy.astro.utoronto.ca} 
lists all variable events targeted by the real-time pipeline,
as well as information returned by spectroscopy.
Snapshots of these databases were extracted in January, 2009 and used 
to match the selected SN-like events of our photometric analysis 
with SNLS real-time candidates. 
Spectral data used for this matching
(\cite{bib:howell},~\cite{bib:bronder},~\cite{bib:balland} and~\cite{bib:walker})
are identical to that used to define the SNLS 3-year sample for
cosmological analyses~\citep{bib:guy2010,bib:conley2010}.

\begin{figure}[htbp]
\begin{center}
\begin{tabular}{cc}
\epsfig{figure=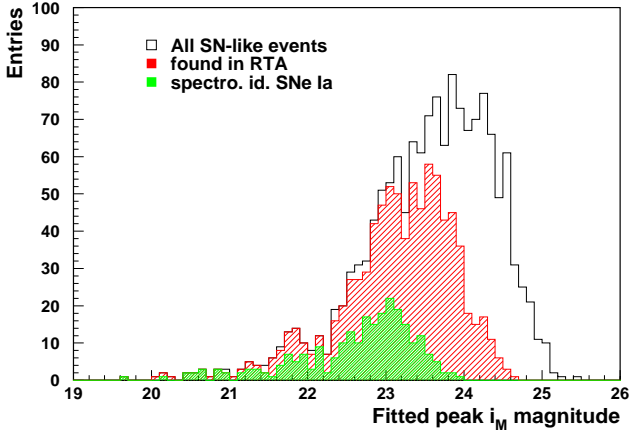,width=\columnwidth}
\end{tabular}
\caption[]{
Fitted $i_M$  peak magnitude distribution of selected SN-like events. 
Among these, events matched to SNLS real-time candidates (resp. 
spectroscopically identified SNe~Ia) correspond to the red
(resp. green) hatched histogram.
} 
\label{fig:db}
\end{center}
\end{figure}

\par Of the 1483 selected SN-like events, 865 (58\%) could be associated
with real-time candidates. 
Most of the remaining 618 events are faint events, 
with $i_M$ peak magnitudes above $23.4$, 
as shown in Figure~\ref{fig:db}.
Besides background events, which are still present at this stage, 
these may also be interesting events too faint to be detected by the real-time
pipeline.
The matched events have the following types in the SNLS databases.
246 events are classified as SNe~Ia and 42 are classified as 
core-collapse SNe, adding in both cases secure and probable events.
These two subsamples are referred to as spectroscopically 
identified events throughout this paper.  
The other events are mainly classified as SNe
(492 potential ones and 52 confirmed as supernovae by spectroscopy). 
A few AGN's (12 potential ones and 3 confirmed events)
and possible stars (11 events, none confirmed) are also present.
The type of these events is mostly based on real-time photometry,
as most of them were not sent for spectroscopy. The break down of
the SNLS database event types in the selected SN sample
is presented in Table~\ref{tab:types}.

\par 
Spectroscopically identified supernovae can be used 
to check the efficiency of the SN selection on bright supernova events.
The SNLS database lists a total of 280 spectroscopically identified SNe Ia and
55 spectroscopically identified core-collapse events in the data covered
by our photometric analysis. All of them are detected by our
pipeline, except for two SNe Ia in D3, one outside our
reference image and one from the presurvey phase.
Of the 333 detected events, 288 passed the SN selection, 
as shown in Table~\ref{tab:presel}.
Most of the lost events were rejected by the sampling cuts. Those were
not fulfilled due to either bad weather conditions or observations during the
presurvey phase (42\%), 
or truncation of the light curve at one end of the observation season (29\%) 
or photometric points missing near maximum light because of 
CCD failures or full moon (24\%).

\subsection{SN selection efficiency for Type Ia supernovae}\label{sec:lceff}
\par The efficiency of the SN selection  was studied further with our 
set of synthetic SN~Ia light curves. Those
were simulated in the observed bands of MegaCam by the SALT2 package. 
Details about the simulation are given
in the section below, comparison with data and selection efficiency
are discussed in the next two sections.

\subsubsection{SN~Ia light curve simulation}\label{sec:lcsimul}
\begin{figure}[htbp]
\begin{center}
\begin{tabular}{cc}
\epsfig{figure=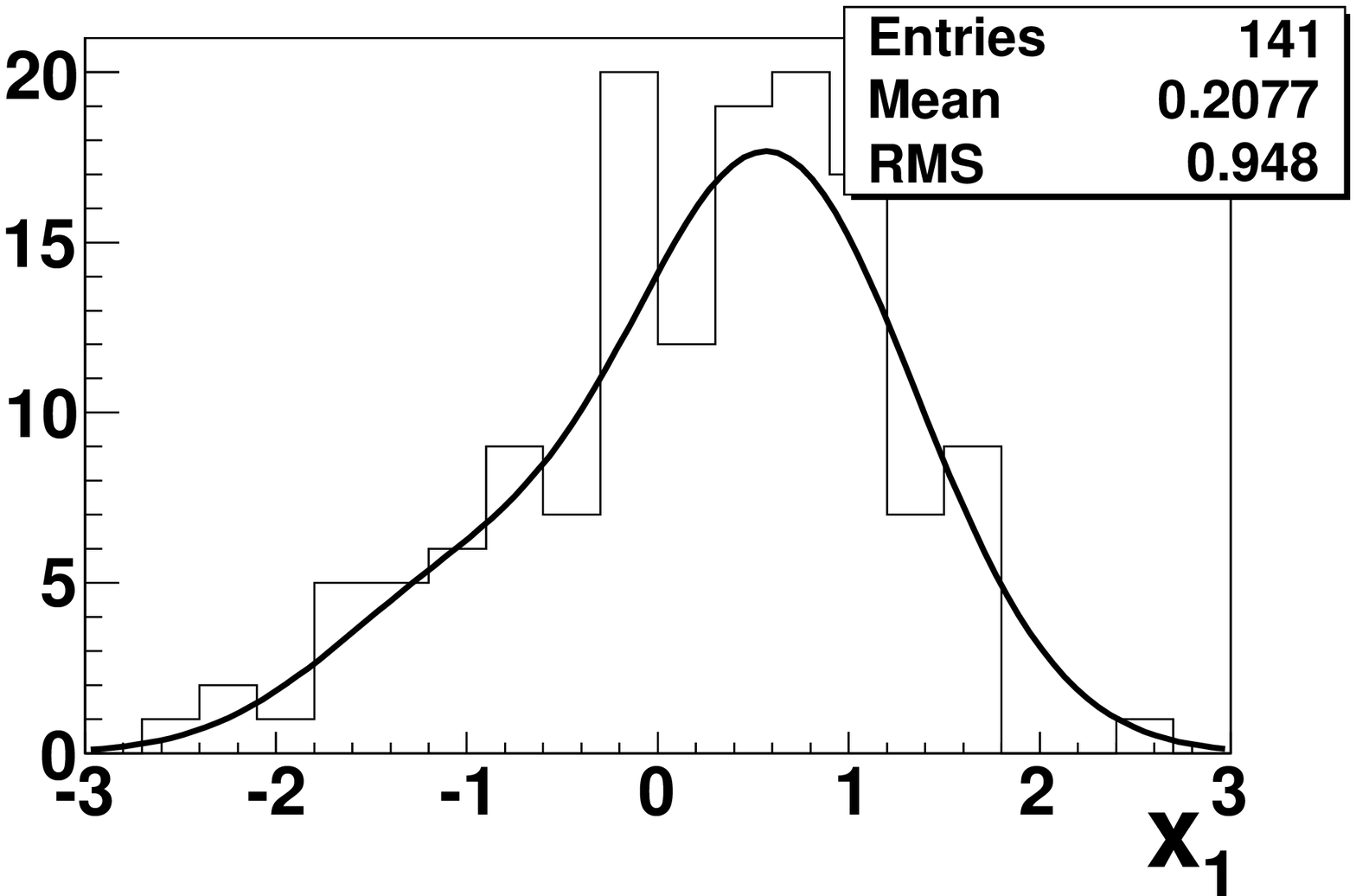,width=0.5\columnwidth}
\epsfig{figure=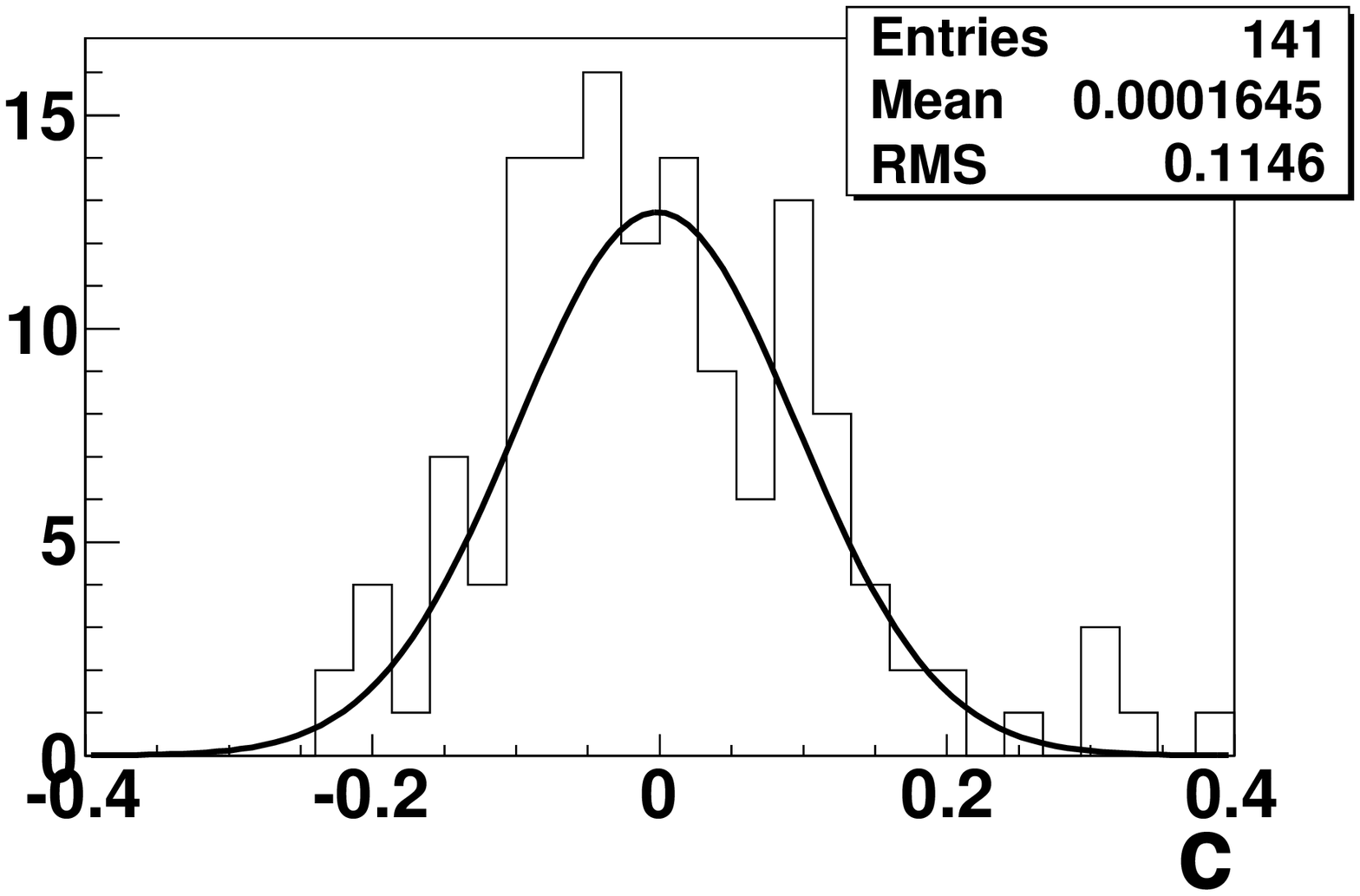,width=0.5\columnwidth}
\end{tabular}
\caption[]{
$X_1$ and colour distributions of spectroscopically
identified SNe~Ia with redshift $z<0.7$. The curves are
used to produce synthetic light curves from the
observed distributions.
} 
\label{fig:input}
\end{center}
\end{figure}

Synthetic SN~Ia light curves were produced assuming a flat $\Lambda$CDM cosmology 
with $\Omega_M = 0.23$.
Redshifts were simulated in the range from 0 to 1.2, with distances 
computed in the above cosmology.
A volumetric distribution was assumed in order to reproduce the acceptance of the 
survey. The redshift dependence of the SN explosion rate is irrelevant for
efficiency studies and was thus ignored at this stage.
Values of the SALT2 $X_1$ and colour parameters were
generated according to model distributions fitted to the 246
spectroscopically identified SNe~Ia present in our selection
(see Table~\ref{tab:presel}),
restricting to redshifts below 0.7 to avoid selection biases.
A Gaussian distribution was used for the colour parameter, with mean 
and $\sigma$ values of 0. and 0.1 (see Figure~\ref{fig:input}).
For $X_1$, a double Gaussian was found more appropriate, with mean 
values of 0.06 and -1.07, $\sigma$ values of 0.8 and 0.7 and relative 
amplitudes equal to 0.25.

For each generated event, a rest-frame $B$-band peak magnitude, 
$m_B^*$, was computed as a function of the generated values 
of $X_1$, $C$ and of the luminosity distance at the generated
redshift in the assumed cosmology, $d_{L}(z,\Omega_M)$ as:
\[ m_B^* = M - \alpha X_1 + \beta C + 5 {\rm log_{10}} d_{L}(z,\Omega_M)\]
using $M= -19.09$, $\alpha= 0.13$ and $\beta= 2.56$.
A Gaussian spread with
$\sigma=0.14$ was applied to $m_B^*$ in order to allow for
SN~Ia intrinsic dispersion.
The magnitude
after dispersion was converted into a global normalisation factor
$X_0$ according to the relationship between $m_B^*$, $X_0$, $C$ and
$X_1$ measured in the spectroscopically identified SNe~Ia.
Finally, dates of $B$-band maximum light were drawn uniformly in intervals starting 
10 days before the beginning of the true observation periods and ending
5 days after their end. 
Each generated SN was also attributed a random position in one of
the four fields of observations.
Corrections for the Milky Way extinction of the SN~Ia flux
at the generated coordinates were applied in SALT2 using the 
dust maps of~\cite{bib:schlegel} and the extinction law 
of~\cite{bib:cardelli}. 

\par Detection and instrumental effects were then simulated in the
following way. The detection efficiency model described in 
Section~\ref{sec:detec} was used to compute the detection probability
of each synthetic event.
Based on this probability, generated events
were then randomly eliminated or kept for further processing.
For each event passing this step, fluxes were computed at dates
corresponding to true observations at the generated SN position in the
mosaic. This allowed us to account for the observing strategy 
as well as for periods of lack of observations due to bad weather 
conditions or temporary CCD inefficiencies.

\par From these ideal fluxes, reconstructed fluxes and errors
were then simulated according to reference distributions determined 
from the 278 spectroscopically identified SNe~Ia that we 
detected (see Table~\ref{tab:presel}).
Distributions were those of flux errors and differences between the measured and 
the SALT2-estimated fluxes (thereafter called shifts in flux). Correlations 
between flux errors, flux values and flux shifts were taken into account 
when building the reference distributions. 
As around 20\% of the light curves in the test-sample 
exhibit more points with a large shift in flux than 
expected from a pure random distribution, the reference 
shift distributions were determined separately for good and bad 
quality light curves.
Observation dates showing systematic large shifts and errors 
were also described by reference distributions determined separately.
A total of 20,000 synthetic light curves was generated.

\subsubsection{Comparison with data}\label{sec:simcompa}

\par
\begin{figure}[htbp]
\begin{center}
\epsfig{figure=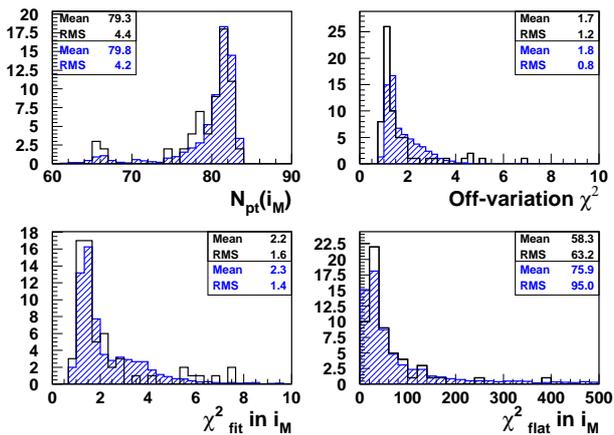,width=\columnwidth}
\caption[]{
Comparison between spectroscopically identified SN~Ia
data (open histogram) and 
synthetic light curves of bright SNe~Ia (hatched histogram) in the D1 field.
Distributions show the total number of points in the $i_M$ light 
curve and three  $\chi^2$ variables used in the analysis (see text).
The mean and RMS of each distributions are indicated in the upper
box for for data and in the lower one for simulation.
} 
\label{fig:simul}
\end{center}
\end{figure}

 Synthetic light curves are compared with spectroscopically identified 
SN~Ia data in Figure~\ref{fig:simul}, where the first step of the
selections described in Section~\ref{sec:preselcuts} and 
an observed peak $i_M$-band magnitude below 23.5 were required.
The agreement is reasonable. In particular, the simulation reproduces
both the peak and the tail of the distributions of the various 
$\chi^2$ variables used in the selections.

The comparison is extended to all data in Figure~\ref{fig:presel2},
showing the qualitative agreement between synthetic SN~Ia 
light curves and identified SN data.
In particular, there are two features that are reproduced by the simulation. 
The $\chi^2$ of the analytical fit, $\chi^2_{fit}$, deteriorates at large 
$\chi^2$ differences, that is at low redshift, where formula~(\ref{fitformula}) 
is less suited to fit SN light curves due to a secondary maximum in
the near-infrared rest-frame light curves that SNLS observes in the $i_M$-band
at low redshift. 
Second, at moderate
$\Delta\chi^2$, the $\chi^2_{fit}$ distribution 
has a tail at values above 2 which, in the simulation,
is reproduced by bad quality light curves, i.e. light curves with a 
sizeable number of photometric points departing from
the generated fluxes (see Section~\ref{sec:lcsimul}).

\begin{figure}[htbp]
\begin{center}
\epsfig{figure=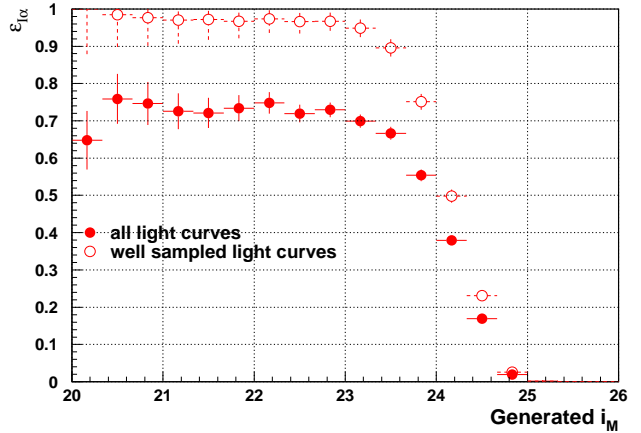, width = \columnwidth}
\caption[]{
Efficiency of the SN selection for Type Ia supernovae as a function of the 
peak magnitude in $i_M$-band, for all events (full dots) and
for events with well sampled light curves (open dots).
The efficiency was evaluated with synthetic SN~Ia light curves.}
\label{fig:preseleff}
\end{center}
\end{figure}

\subsubsection{Efficiency}\label{sec:preseleff}

\par The selection efficiency for Type Ia supernovae is presented in 
Figure~\ref{fig:preseleff} as a function of the generated peak magnitude in 
$i_M$-band (full dots). The four fields were combined 
and, as fiducial cuts, we restricted all efficiency computations 
to those events whose generated date of $B$-band maximum light was strictly
within the observing periods.
For comparison, Figure~\ref{fig:preseleff} gives also the efficiency 
for well sampled light curves (i.e. events passing the sampling cuts 
of the selection), showing that, at this stage of the analysis, the 
efficiency loss for bright events is due mostly to sampling requirements.

The efficiency for bright synthetic events is detailed in the last column of 
Table~\ref{tab:presel}, which also gives that for spectroscopically 
identified SNe~Ia in data.
The efficiency from simulation ($0.73/0.96=76\%$) is lower than that for
data events ($246/278=88\%$) since these have been
submitted to tighter quality criteria to ensure conclusive spectroscopic
measurements. As an example, bright events with a light curve 
starting past maximum light were not sent for spectroscopy, while they are 
included in the simulated sample. 
Restricting again the comparison to well sampled light curves in both 
simulation and data, the efficiencies become $246/246=100\%$ in data and 
$0.97/0.98=99\%$ in simulation, which gives confidence  that
reliable selection efficiencies are deduced from the simulation.

\subsection{Redshift assignment}\label{sec:zassignment}

\par The selected SN-like events were matched with galaxies in the catalogue 
described in Section~\ref{sec:ctlg}.
The match was considered successful if the nearest galaxy from an
event was found within a distance of $5r_{\rm gal}$, where the galaxy's
effective radius, $r_{\rm gal}$, was defined as the half-width of the galaxy
in the direction of the event and derived from the A,B and $\theta$ 
SExtractor parameters. 
In this procedure, galaxies so faint that no size measurement could be 
performed (typically fainter than $i_M \sim 24.5 $) 
were assigned a size of 1~pixel 
and were considered only if no bright galaxy was found in the vicinity of
the event. 
The choice of the $5r_{\rm gal}$ threshold was a compromise between host
finding efficiency and accidental mismatching.
Of the 1483 selected SN-like events, 1352 (91\%) had matched hosts
and 1233 of these (91\%) had a photometric redshift which was then 
attributed to the event. 
The types of these events as found in the real-time database of SNLS
(see Section~\ref{sec:lcrta}) are listed in Table~\ref{tab:types}.


\par
\begin{figure}[htbp]
\begin{center}
\epsfig{figure=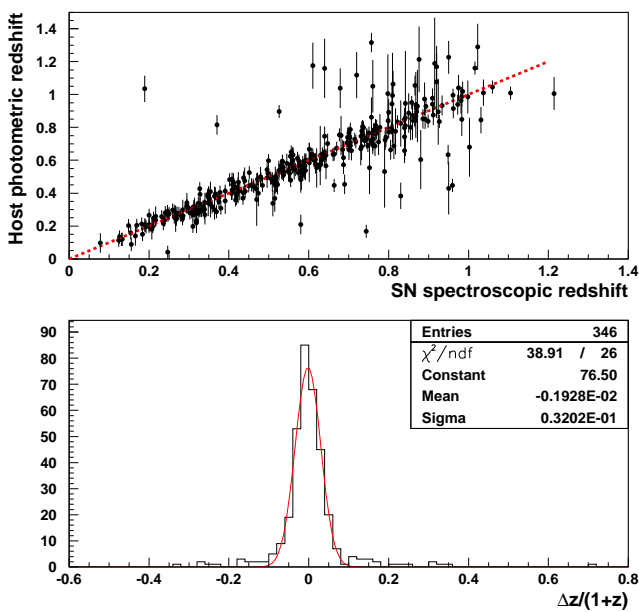, width =  \columnwidth}
\caption[]
{Host galaxy photometric redshifts compared with event spectroscopic redshifts in
the subsample of events with both measurements available at the end of the SN selection.
The resolution of galaxy photometric redshift measurements is 
given in the bottom plot.} 
\label{fig:preselz}
\end{center}
\end{figure}

 Host galaxy photometric redshifts were compared to event spectroscopic
redshifts in the subsample of 346 selected SN-like events which have both 
measurements. The result is shown in Figure~\ref{fig:preselz}. 
In most cases, the two redshift measurements agree and
the galaxy photometric redshifts provide a resolution of 
$\sigma_{\Delta z/(1+z)} \sim 3\%$ in the central part
of the distribution.
However, 5.2\% (resp. 10\%) of the galaxy photometric redshifts 
depart from the spectroscopic redshifts 
by more than 5 (resp. 3) $\sigma$. 
Such differences can be explained by incorrect SN-to-galaxy 
associations or by bad galaxy photometric redshifts. 
The latter effect is likely to be the main reason for the
deviations. According to~\cite{bib:ilbert2006}
3.7\% of the galaxy photometric measurements are
to be expected with $\Delta z / (1+z) > 0.15$, for
a sample of galaxies with magnitudes below 24 in 
$i_M^{AB}$, which corresponds to the
range of host galaxy magnitudes in this study.

\par Incorrect redshift assignments are likely to generate significant
systematic biases in the subsequent photometric SN~Ia selection, 
especially at large ($z>0.6$) redshifts, where most of the incorrect assignments
occur (see Figure~\ref{fig:preselz}). Special care was thus devoted
to identify and reject most of these events in the SN~Ia selection criteria
that are described in the next section.
As synthetic light curves provide an important tool to define the selections,
host galaxy photometric redshifts were also simulated for the sample of synthetic 
light curves. 
Redshifts were generated according to a Gaussian distribution in 
$\Delta z / (1+z)^2$, to account for the fact that 
the reliability of the galaxy photometric redshifts decreases significantly 
for fainter objects~\citep{bib:ilbert2006}.
Redshift outliers were generated with a Gaussian distribution in 
$\Delta z / (1+z)$ and assuming their rate rises linearly with redshift.
Parameters for this simulation were derived from the comparison between
host galaxy photometric redshifts and event spectroscopic redshifts in the data, 
using the test-sample of 346 events already mentioned. 
Finally, the fraction of simulated events with a host redshift after the SN 
selection was given the same probability 
as observed in the data.

\begin{table}[htb]
\caption[]{Event type break down at different analysis stages.}
\label{tab:types}
\begin{center}
\begin{tabular}{lrrr} \hline \hline
Event         & \multicolumn{3}{c}{Analysis step}\\
class         & SN sel. & +$z_{gal}$  & Ia sel.\\ \hline
All events    & 1483  &  1233   & 485 \\\hline
In database   &  865  &   739   & 388 \\\hline
Ia type       &  246  &   208   & 175 \\
CC types      &   42  &    38   & 1$^{*}$\\
SN-like       &  544  &   467   & 210\\
AGN-like      &   15  &    13   & 2\\
Variable-like &   11  &     8   & 0\\
Unknown type  &    7  &     5   & 0\\
\hline
\end{tabular}
\tablefoot{
The break down 
is given in the photometrically selected SN-like events (column 2), 
in those events with a host galaxy photometric redshift (column 3)
and in those finally kept as SN~Ia events (column 4).
The type of the event marked with an asterisk was revisited
after this analysis (see Section~\ref{sec:compa_rta}).
} 
\end{center}
\end{table}

\subsection{Summary}
The total sample of 1233 SN-like events with a host galaxy 
photometric redshift is the starting point of the SN~Ia search described
in the next section. 
A measurement of the rate of core-collapse supernovae at low redshift 
was also deduced from that sample, as described 
in~\cite{bib:ratecc}.~\footnote{This paper used a 
preliminary flux calibration resulting in slight differences in the 
analysis (e.g. 1207 SN-like events with a host galaxy photometric redshift 
instead of 1233)}

\section{SN~Ia selection}\label{sec:sn}
The photometric SN~Ia selection cuts were designed to discriminate SNe~Ia
from core-collapse supernovae (SNe~CC) 
and to assess the reliability of the redshift assignment.
The search was restricted to normal SNe~Ia only, and relied on the SALT2
light curve fitter.
As for the SN selection, the criteria were set up using  
spectroscopically identified supernovae present in our
sample and synthetic SN~Ia light curves as qualitative guidelines.  

\subsection{Selection criteria}\label{sec:sncuts}

\par The four-band light curves of each event were fit simultaneously 
with SALT2, fixing the redshift to the host galaxy photometric redshift, 
to derive the date of $B$-band maximum light $t^{B}_{max}$,
the colour $C$, the stretch-related parameter $X_1$ and the rest-frame  
$B$-band peak magnitude $m_B^*$ under the assumption that the event 
was a SN~Ia. The fit converged for 98\% of the events. 

\subsubsection{Light curve sampling}
To ensure that meaningful SALT2 parameters were obtained, tighter quality cuts 
were applied w.r.t. the SN selection.
Defining the event rest-frame phase as $\tau=(t_{obs}-t^{B}_{max})/(1+z_{gal})$,
events were required to fulfil the following minimum conditions~\footnote{
These requirements are very similar to those used to define the SNLS 3-year 
cosmological samples~\citep{bib:guy2010}}~: 
\begin{itemize}
\item[(i)] at least one measurement in the range $-10<\tau<+5$ days for a 
reasonable estimate of $t_{max}$ and thus of the peak magnitude, 
\item[(ii)] at least one measurement in the range $+5<\tau<+20$ days for 
a reasonable shape evaluation, 
\item[(iii)] at least one colour among ($g-i$), ($r-z$) and ($i-z$)
with at least one measurement in each band in the range $-10<\tau<+35$ 
for a reasonable Ia/CC discrimination (see Section~\ref{sec:cmdiag}). 
\end{itemize}
After these cuts,
the selected sample contained 1152 events, among which 203
(resp. 35) were confirmed as SNe~Ia (resp. SNe~CC) by spectroscopy. The
remaining selections intended to enrich the sample in actual SN~Ia events
with correct assigned host galaxy photometric redshifts.

\subsubsection{$\chi^2$ requirements}\label{sec:chi2}
To reject SN~Ia light curves with a poor fit and to start eliminating non-Ia SNe, 
cuts were applied to the $\chi^2$ per degree of freedom in each filter,  
$\chi^2<10$ in the $g$ filter and $\chi^2<8$ in the $r,i,z$ filters. 
In addition,
the $\chi^2$ per degree of freedom of the overall fit was required to be less than 6.
Seven spectroscopically confirmed SNe Ia were rejected by these conditions. 
Two were assigned an incorrect host photometric redshift, the other five 
exhibited noisy light curves.
These cuts also discarded nine spectroscopically confirmed SNe~CC, all 
of the SN~II type.

\par The light curve fit with SALT2 only included 
MegaCam filters corresponding to the redshifted wavelength interval between 
290~nm and 700~nm, where the SN~Ia model is defined with sufficient accuracy 
(see~\cite{bib:guy2007} for details). Filter $g_M$ was thus only used below a 
redshift of about $0.68$, while $z_M$ was considered only for a redshift larger than 
$0.26$.  
Outside these redshift intervals, SNe~Ia are expected to give no significant signal 
in these bands, but other types of supernovae may have a different behaviour.
Comparing the
observed flux in the unfitted bands to that corresponding to the SALT2 fit 
can give a complementary constraint against core-collapse events, as well as
against incorrect redshift assignments in true SNe~Ia.

\par The $\chi^2$ per degree of freedom with respect to the SALT2 best fit was thus 
computed in the unfitted bands and required to be less than 6 in $g_M$, and less than 
3 in $z_M$. 
This criterion is stronger than that for the bands entering the fit, to compensate 
for the fact that the subsequent selections are
looser for events which do not have all bands included in the fits.
Ten spectroscopically confirmed SN~Ia were eliminated by this second set of cuts. 
They were all assigned a bad host photometric redshift, with 
$\langle |z_{gal}-z_{spe}|\rangle$ = 0.35 compared to an average of 0.06 for 
the other confirmed SNe~Ia. 
These cuts also rejected four spectroscopically confirmed core-collapse events 
(two SNe~II, two SNe~Ib/c) because of their high $\chi^2$ in $z_M$.

\subsubsection{Stretch and colour constraints}
\par
\begin{figure}[htbp]
\begin{center}
\epsfig{figure=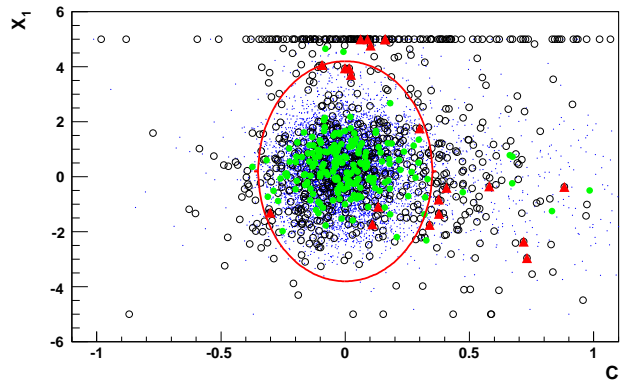,width=\columnwidth}
\caption[]{$X_1$ vs. $C$ for synthetic SNe~Ia (blue dots) and  
for data (all other symbols), after sampling and $\chi^2$ constraints 
have been applied.
Green filled circles and red triangles stand for spectroscopically identified 
SNe~Ia and SNe~CC present in our sample, respectively. 
Open black circles stand for data events with no spectroscopic identification.
Events outside the red ellipse are rejected. The pileup at $X_1$=5 is
due to an upper bound encoded in the fitter.}
\label{fig:cgal2}\end{center} 
\end{figure}

 
As a further discrimination between SN~Ia and non SN~Ia events, a constraint was 
applied on the fitted $X_1$ and colour parameters, requiring 
\[\left(\frac{X_1 - 0.2}{4.0}\right)^2 + \left(\frac{C}{0.35}\right)^2 < 1 \;\; \]
as illustrated in Figure~\ref{fig:cgal2}. 
The constraint accounts for the offset of 0.2 observed in the stretch 
distribution, as shown in Figure~\ref{fig:input}. 
Such an offset may reflect the higher number of high stretch SNe~Ia observed at high 
redshift due to the increased fraction of star forming host galaxies in the past
universe, which lead to more luminous supernovae~\citep{bib:howell07,bib:sullivan10}.

While the lower bound on $X_1$ only removed a few relatively short and  
faint (average peak magnitude in $i_M$ of 25) data events, the upper bound  
discarded a large number of long duration events among which 
seven spectroscopically confirmed SNe~CC, all of SN II type.
On the other hand, the constraint on $X_1$ removed a 
negligible fraction (1\%) of the synthetic SNe~Ia.

\par The constraint on the fitted colour relies primarily on the observation that, 
while SN~Ia colours were centred around 0 (with a r.m.s. of 0.11), 
the spectroscopically identified core-collapse events in our sample exhibited 
an average colour $\langle C \rangle=0.3$. Requiring $C<0.35$ rejected
nine of these, mostly SN~Ib and SN~Ic events.

\par
\begin{figure}[htbp]
\begin{center}
\epsfig{figure=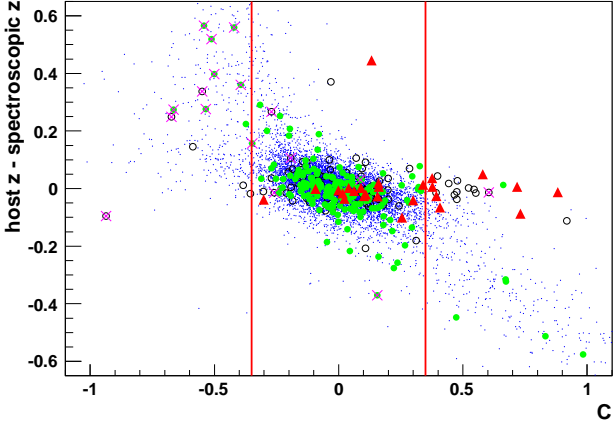,width=\columnwidth}
\caption[]{Difference between redshift assignments as a function of SALT2
fitted colour. Same colour code and level of selections as in Fig.~\ref{fig:cgal2},
except for pink crosses which indicate data events rejected by the $\chi^2$ 
selections in unfitted bands. 
Red lines indicate the  extreme values of the colour cuts.
}
\label{fig:cgal}\end{center} 
\end{figure}

 The cut on colour also allowed the rejection of SNe~Ia where the associated 
host galaxy photometric redshift, $z_{gal}$, was far from the true one. As
an illustration, Fig.~\ref{fig:cgal} presents the difference between $z_{gal}$ 
and the more precise spectroscopic redshift, $z_{spe}$ as a function of 
the fitted event colour, for events with both redshift assignments.
Extreme colours in identified SNe~Ia are mostly associated with inaccurate 
galaxy photometric redshift measurements. 
Synthetic SNe~Ia show the same trend. The accuracy with which the colour of 
a true SN~Ia is recovered, is indeed directly related to the reliability of 
the redshift assignment.
As an example, about 70\% of the synthetic light curves with 
$\left| z_{gal}-z_{spe} \right| > 0.2$ have $|C|$ above 0.35. 


\subsubsection{SALT2 colour-magnitude diagrams}\label{sec:cmdiag}
\par
\begin{figure}[htbp]
\begin{center}
\epsfig{figure=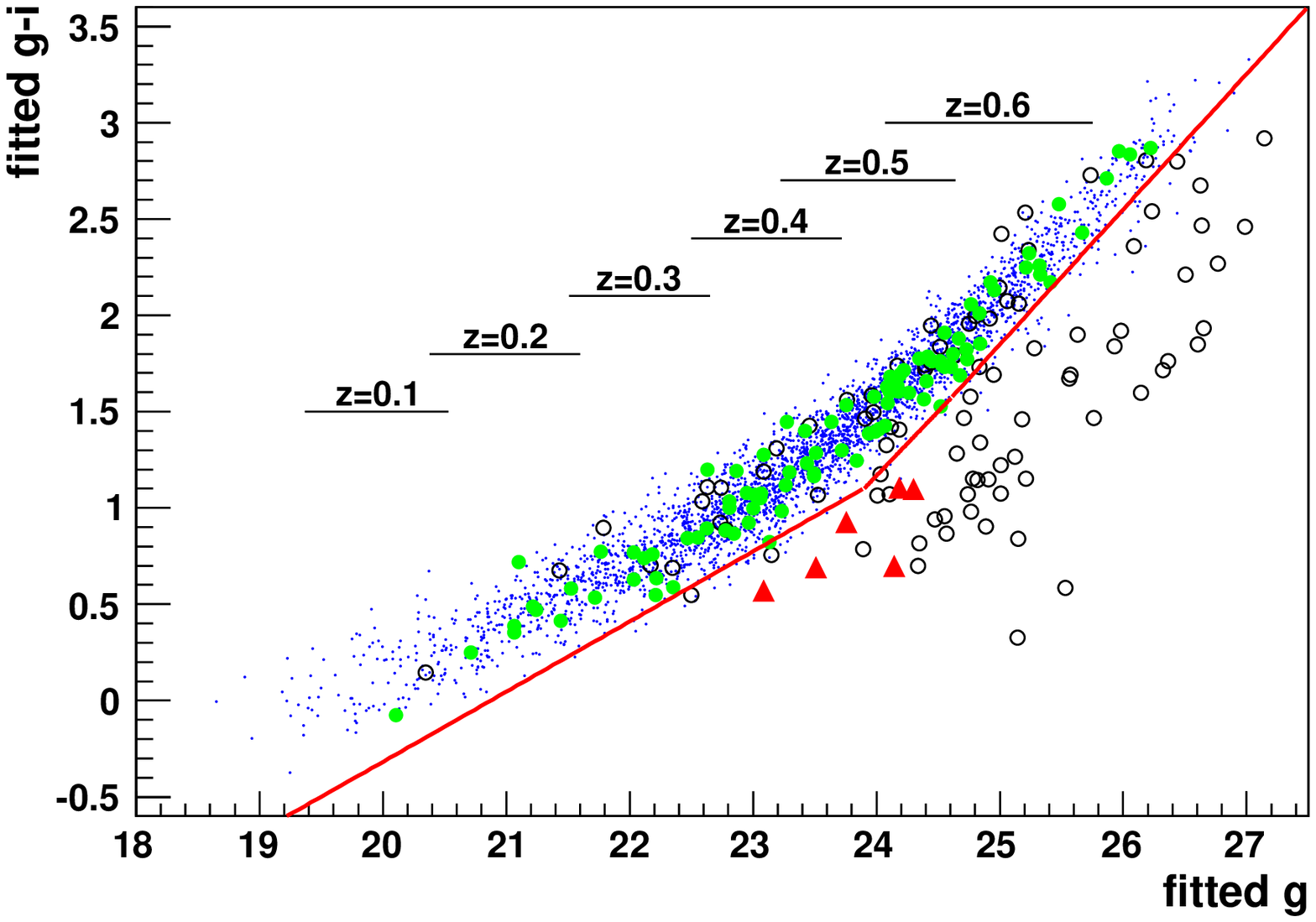,width=\columnwidth}
\vskip -0.2cm 
\epsfig{figure=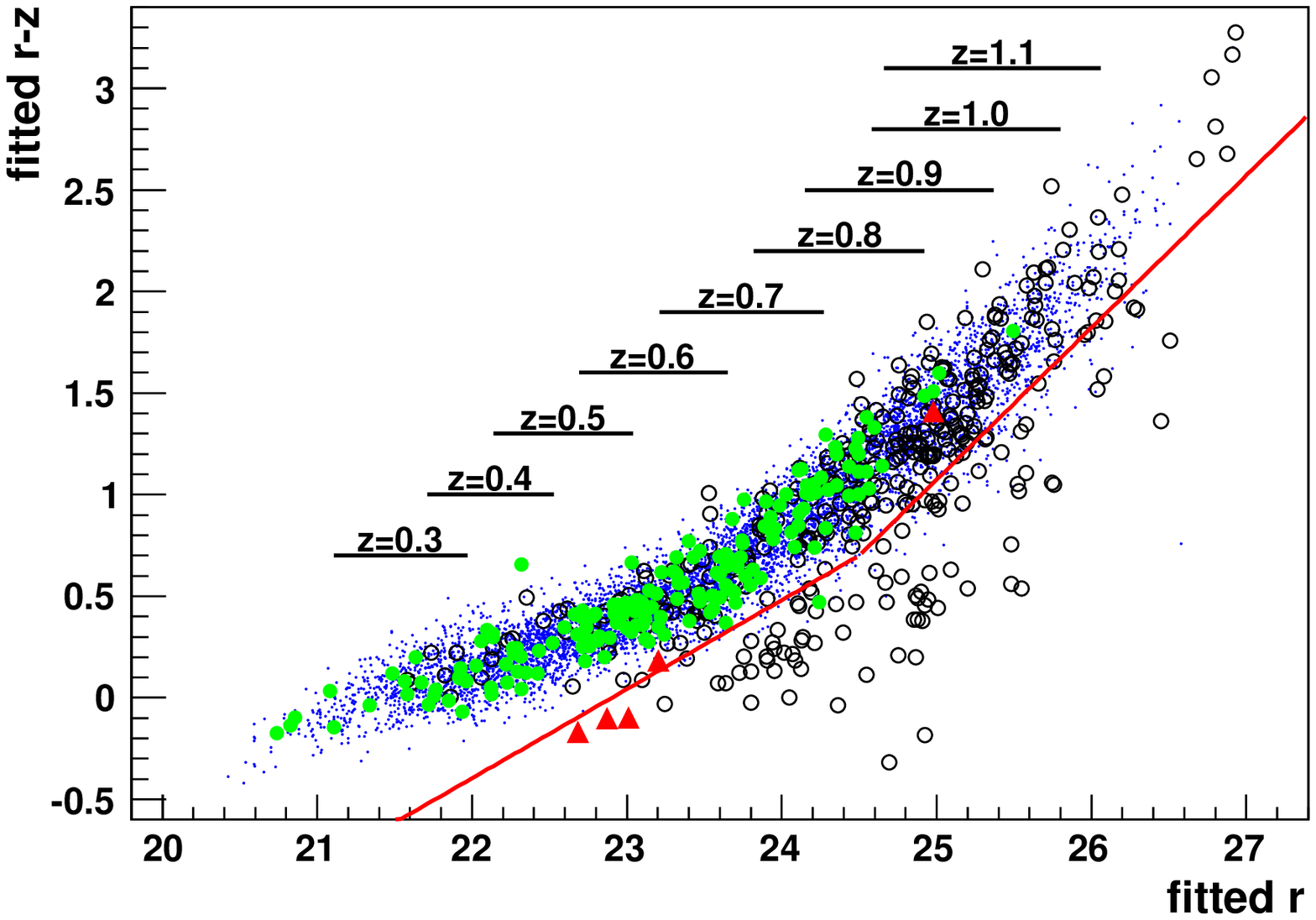,width=\columnwidth}
\vskip -0.2cm
\epsfig{figure=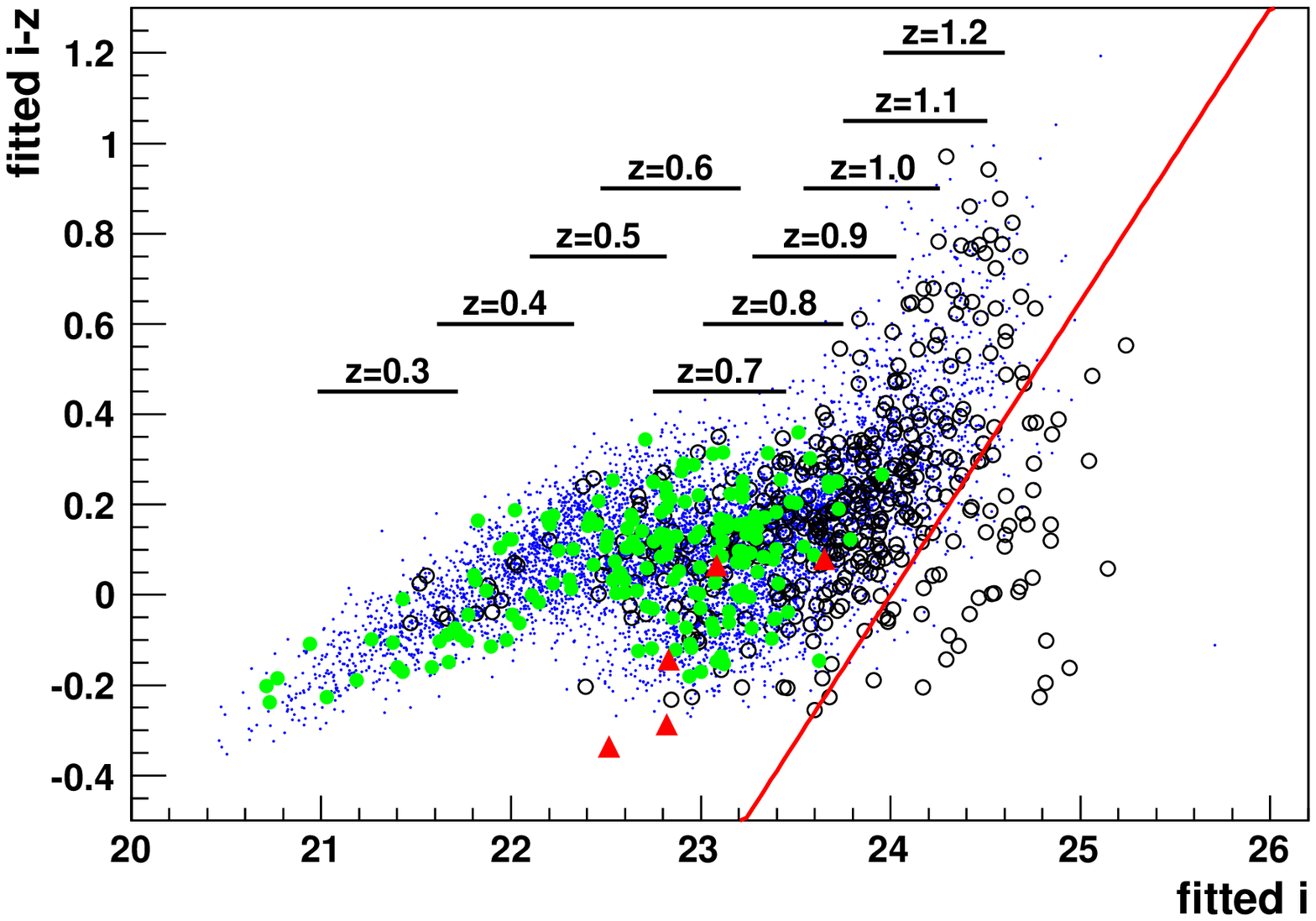,width=\columnwidth}
\caption[]{Colour-magnitude diagrams based on SALT2 fitted 
magnitudes for events that passed all previous constraints up to 
$X_1$ and $C$ cuts. Same colour code as in Fig.~\ref{fig:cgal2}.
Events below the lines were rejected. $1\sigma$ magnitude ranges are indicated
for 0.1 redshift bins centred on the indicated redshift values (vertical
locations of the segments are arbitrary).
}
\label{fig:gi}\end{center} 
\end{figure}

 The last criteria to reject non SN~Ia events were based on 
colour-magnitude diagrams from SALT2 fitted magnitudes. In these diagrams,
Type Ia supernovae populate a thin band while core-collapse supernovae 
lie in a broad region which is shifted w.r.t. the SN~Ia band.
To emphasize the colour and magnitude difference between the two classes
of events, the following diagrams were chosen: 
$g-i$ vs. $g$,  
$r-z$ vs. $r$ and
$i-z$ vs. $z$ (Fig.~\ref{fig:gi}, top to bottom),
where $g$, $r$, $i$ and $z$ are the magnitudes in the MegaCam filters computed 
from the SALT2 best fit model, at the date of $B$-band maximum light.
Note that events entered these diagrams only when the filters of interest 
were actually considered in the fit (see Sec.~\ref{sec:chi2}). The first 
diagram was thus used for low redshift events, $0.16<z<0.68$, the second 
one was considered in the redshift range $0.26<z<1.15$, while the third one 
applied for $0.26<z$.
Events at the highest redshift ($z>1.15$) were thus tested only through 
the third diagram.

\par The $g-i$ vs. $g$ diagram proved to be very helpful in reducing the 
remaining background from core-collapse supernovae, as most of those we 
could detect were at low redshift, and, in this diagram,
appeared fainter than actual SNe~Ia for the same  $g-i$ colour. 
Although the magnitude dispersion of the SNe CC is close to 1 mag, this 
diagram offered a good discrimination between CC and Type Ia 
supernovae. Part of the SNe~CC were detected at high enough 
redshift to enter the $r-z$ vs. $r$ or $i-z$ vs. $i$ diagrams.
A similar trend was observed there, with a population distinct from 
the SNe~Ia, lying at fainter magnitudes for a given colour. 
The separation between the two populations in these diagrams is however 
less pronounced than in the $g-i$ vs. $g$ diagram. 

Altogether, the three diagrams were useful to complete the reduction of the CC 
supernova background.
Note however that the spectroscopically confirmed SNe CC that 
would be rejected by the colour-magnitude diagrams alone,
i.e. not considering any of the previous SN~Ia selection criteria,
are all typed as either SN-II or SN-IIP. In the colour-magnitude diagrams,
the SNe~Ib/c often lay close to the upper limit of the SN~Ia band and 
could not be identified unambiguously.

\subsubsection{Results}

\par
\begin{figure}[hbtp]
\begin{center}
\epsfig{figure=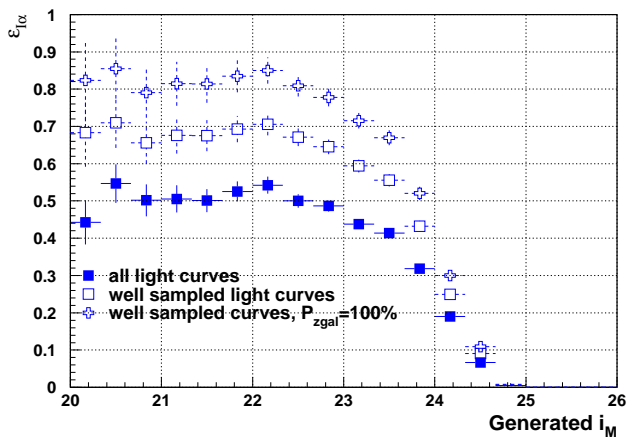, width = \columnwidth}
\caption[]{
Efficiency of the SN~Ia selection for Type Ia supernovae
as a function of the peak
magnitude in $i_M$-band, for all events (full squares) and
for events with well sampled light curves, when the host
galaxy redshift assignment probability is included
(open squares) or not (crosses).
The efficiency was evaluated with synthetic SN~Ia light curves.}
\label{fig:seleff}
\end{center}
\end{figure}

 At the end of the selection, 485 SN~Ia candidates were selected. 
The overall selection efficiency is given in Figure~\ref{fig:seleff} 
as a function of the peak magnitude in the $i_M$ filter, 
for all synthetic events (full squares) and for events with well sampled 
light curves (open squares) defined as in Section~\ref{sec:preseleff}.
For the latter, the bright event efficiency is 67\%. It would be 
80\% if all events were assigned a host galaxy photometric redshift.
This number represents the intrinsic performance of our selection for
bright SNe~Ia, regardless of the survey observing strategy or the
availability of host redshifts. 

The effect of the different selection criteria is detailed in 
Table~\ref{tab:selIa}, where the selection efficiency is given for bright
events with well sampled light curves.
As for the SN selection, the efficiency of the SN~Ia selection criteria from 
simulation ($0.67/0.97=69\%$) agrees well with that for spectroscopically 
identified SNe~Ia ($175/246=71 \pm 3\%$).

\begin{table}[htb]
\caption[]{Effect of the SN~Ia selection cuts.}
\label{tab:selIa}
\begin{center}
\begin{tabular}{lrrrr} \hline \hline
Cut & Events & Ia & CC & $\epsilon$(Ia) \\ \hline
SN selection               & 1483 & 246 & 42 & 0.97 \\
Host redshift available    & 1233 & 208 & 38 & 0.80 \\
Light curve sampling       & 1152 & 203 & 35 & 0.75 \\
$\chi^2$                   & 951 & 186 & 22 & 0.68 \\
$X_1$ and colour         & 596 & 176 & 7 & 0.67 \\
$g-i$ vs. $g$              & 539 & 176 & 1$^{*}$ & 0.67 \\
$r-z$ vs. $r$              & 490 & 175 & 1$^{*}$ & 0.67 \\
$i-z$ vs. $z$              & 485 & 175 & 1$^{*}$ & 0.67 \\
\hline
\end{tabular}
\tablefoot{The effect of the cuts 
is given for all detections (column 2), for the
subsamples of events identified  by spectroscopy as 
Type Ia and core-collapse SNe (columns 3 and 4), and for
synthetic bright SNe~Ia ($m_{0i}<23$) with well sampled
light curves (column 5). 
The type of the event marked with an asterisk was revisited
after this analysis (see Section~\ref{sec:compa_rta}).
} 
\end{center}
\end{table}

\subsubsection{Discussion}\label{sec:discus}
The use of colour-magnitude diagrams in the selection may generate 
biases in the selection as a function of redshift. To check this, we show
in Figure~\ref{fig:effIa} the selection efficiency as a function of the 
generated redshift (see Appendix~B for the colour and $X_1$ variations).
The dependence seen in Figure~\ref{fig:effIa} follows mostly from the fact that 
the selection is easier for brighter SNe~Ia, i.e. for lower-redshift, bluer and 
higher-stretch events.
The  SN~Ia selection efficiency starts to decrease at redshifts above 0.6 while
the SN selection efficiency plateau extended up to $z \sim 0.8$.
We checked that this decrease is due to the cuts on colour and on the reduced $\chi^2$ 
from the unfitted bands, which are essential to reject non-SNIa contaminants and 
true SNe~Ia which were assigned an incorrect redshift.

\begin{figure}[htbp]
\begin{center}
\epsfig{figure=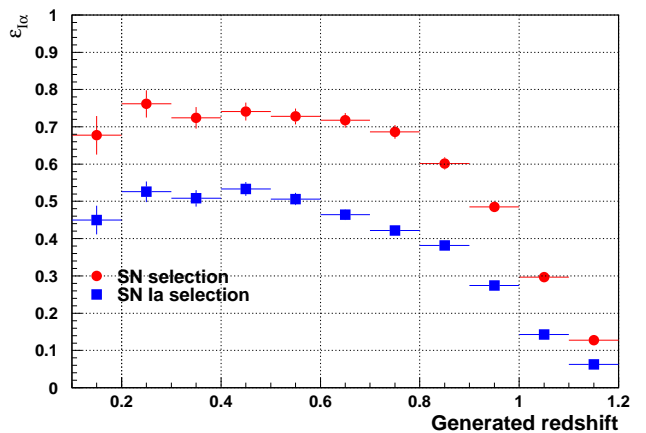, width=\columnwidth} 
\caption[]{
Selection efficiency from synthetic SN~Ia light curves as a function of the 
generated redshift, at different stages of the analysis.} 
\label{fig:effIa}
\end{center}
\end{figure}

The colour-magnitude diagrams used in the above analysis are based on SALT2 
fitted magnitudes and not on colours and magnitudes derived directly from 
observations.
We checked that using magnitudes at the SALT2 date of $B$-band maximum light  
derived from fits to the light curves with formula~(\ref{fitformula}) 
would indeed lead to a poorer discrimination
between SNe~Ia and contaminants, due to larger dispersions, especially in the
$r-z$ vs. $r$ and $i-z$ vs. $i$ diagrams. 
More quantitatively, applying the selections described in Section~\ref{sec:cmdiag} 
on these diagrams would lead to 6\% less efficiency and 
75\% more contamination by CC~SNe (see Section~\ref{sec:conta}).

\begin{figure}[htbp]
\begin{center}
\epsfig{figure=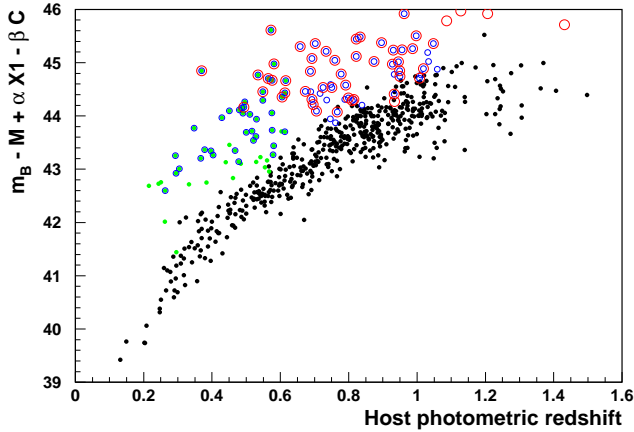,width=\columnwidth}
\caption[]{Hubble diagram for data events at different levels of the 
photometric SN~Ia selection.
Black dots stand for events at the end of the selection, 
green filled (resp. open blue) circles for events rejected by the constraints on the
SALT2 fitted $g-i$ vs $g$ (resp. $i-z$ vs $i$) colour-magnitude diagram, open red
circles for events rejected by the constraints on the
SALT2 fitted $r-z$ vs $r$ diagram. }
\label{fig:hubcuts}\end{center} 
\end{figure}

\par The impact of the colour-magnitude constraints based on SALT2 fitted 
magnitudes is illustrated in Figure~\ref{fig:hubcuts} which shows the Hubble 
diagram of the sample of photometrically selected events 
(see Section~\ref{sec:char} 
for more details about the distance modulus computation) as well as events 
rejected by each constraint. On average, the average distance modulus of the 
rejected events is 1.34 mag larger than that of the selected sample, with a 
r.m.s. of 0.61~mag, almost twice that of the selected sample. 
Keeping events whose distance modulus is within 0.55~mag above the Hubble flow 
would lead to similar SN~Ia efficiency, core-collapse contamination and sample of 
data events as the colour-magnitude diagram constraints used in this analysis. 
The two methods are thus equivalent.

\begin{figure}[htbp]
\begin{center}
\epsfig{figure=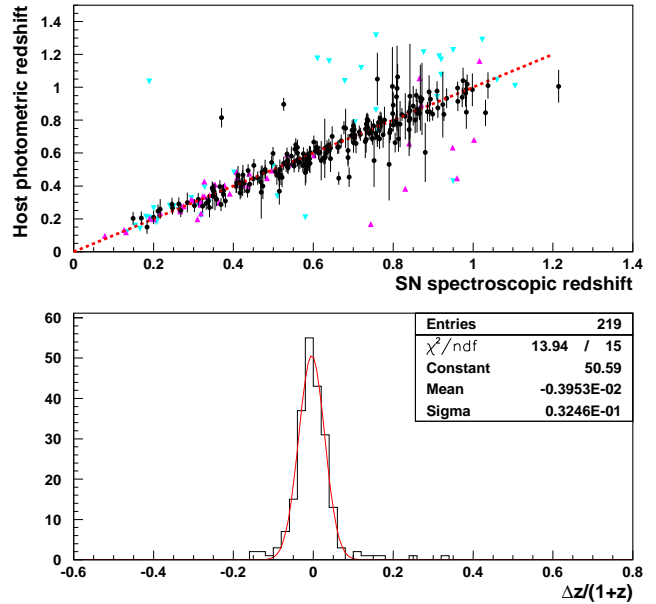, width =  \columnwidth}
\caption[]
{Host galaxy photometric redshifts compared with event spectroscopic redshifts in
the subsample of events with both measurements available after all 
SN~Ia selection criteria have been applied (black dots). 
Downward light blue (resp. upward pink) triangles stand for
SN-like events rejected by the $\chi^2$ 
(resp. colour) cuts of the SN~Ia selection. The redshift resolution 
at the end of the SN~Ia selection is given in the bottom plot.} 
\label{fig:selz}
\end{center}
\end{figure}

\par As mentioned in Section~\ref{sec:zassignment}, 
redshift outliers were present in a significant fraction of the selected SN events,
e.g. 10\% of those had redshifts which were 3 $\sigma$ outliers.
Figure~\ref{fig:selz} shows the comparison between spectroscopic and host galaxy 
photometric redshifts at the end of the SN~Ia selection. 
Highlighted in the plot is the effect of the  $\chi^2$ and colour requirements, 
which prove to be helpful to reject events which were assigned a redshift outlier.
At the end of the SN~Ia selection, 
1.4\% (resp. 5.5\%) of the host galaxy photometric redshifts depart from the 
spectroscopic ones by more than 5 (resp. 3) $\sigma$. The outlier fractions 
at the SN selection level 
have been reduced by a factor 4 and 2, respectively. 
There is no change in the redshift resolution, but the redshift distribution
is now much closer to a Gaussian, as shown in the bottom plot of 
Figure~\ref{fig:selz}.

Summarising, with the above photometric selection, 
well sampled light curves of bright SNe~Ia are selected with an average 
efficiency of 67\%, which includes
the 83\% efficiency of the host galaxy photometric redshift assignment.
Photometric redshifts derived from supernova light curves would offer a better
redshift assignment probability than what we achieved here.
In~\cite{bib:palanque} we set up such a method based on SALT2 and 
tested it with the synthetic SN~Ia light curves and the photometric data sample 
described in the previous sections. Combining both supernova selection
and redshift determination from photometry will be the subject of another 
paper.

\par In the next sections, we pursue the discussion on the present analysis,
comparing its output to that of the real-time SNLS analysis and evaluating the
contamination of the selected sample by residual core-collapse supernovae. 

\subsection{Comparison with the SNLS real-time selection}\label{sec:compa_rta}

As in Section~\ref{sec:lcrta}, events of the photometric sample were
associated with SNLS real-time candidates. The break-down of 
the event types is summarised in Table~\ref{tab:types}.
Out of the 485 candidates, 388 (80\%) were detected in real-time. Of these, 45\% 
are spectroscopically identified SNe~Ia (adding up secure and probable SNe~Ia) and 
another 54\% were typed (mostly photometrically) as SNe. The remaining 3 events
have the following characteristics.

One event, \object{SNLS 06D2bo} was identified as having a spectroscopic redshift of 
z=0.370, on the basis of strong  [OII] and [OIII] emission lines from the 
host galaxy.  The best match appeared to be the broad-lined \object{SN 1998bw}, so 
the event was classified as a possible CC SN. 
The event is highlighted in Figure~\ref{fig:gi} (middle plot)
as the red triangle at $r_M \sim 25$.  
It lands within the locus of points defined by SNe Ia. The photometric 
redshift of the host was much higher, $z=0.82$, and agreed with the estimate 
of the redshift derived from fitting the SN light curve (using 
e.g.~\cite{bib:sullivan} or ~\cite{bib:palanque}). 
The large discrepancy between the photometric redshifts (SN and host) and 
the spectroscopic one prompted us to re-examine the spectroscopic data. 
We soon discovered that the spectrum of an unrelated object had been extracted 
instead of SNLS~06D2bo. The data were reprocessed. 
SNLS~06D2bo is clearly a SN Ia, with a best fit redshift of $z=0.79$. 
We have chosen to let the error in the extraction stand. It vindicates the 
accuracy of the technique we have developed to photometrically identify SNe Ia, 
and it indicates the level of human error in the analysis of the SNLS 
spectroscopic data.The impact of excluding this SN from this paper specifically 
and from the 3-year SNLS analyses more generally is negligible. 

The other two candidates were considered as ``AGN?'' from their real-time 
light curves and were not sent for spectroscopy.
The distances to their host galaxies are 0.2 and 0.3 $r_{\rm gal}$ where
$r_{\rm gal}$ is the effective galaxy radius introduced in 
Section~\ref{sec:zassignment}.
These values are in the bulk of the distance distribution, which peaks near  
0.01 $r_{\rm gal}$. There is no hint that these two events are highly centred, 
as would be expected for AGN's. Their light curves do not exhibit any peculiarity 
either when compared to normal SNe~Ia. 
In both cases, the redshift of the associated host is in agreement with a 
supernova photometric redshift derived from the light-curves.

Finally, SNLS detected eight supernovae which were classified as peculiar Type Ia 
after spectroscopy, as described in~\cite{bib:balland}, \cite{bib:bronder} and 
\cite{bib:ellis}: \object{SNLS 03D3bb}, \object{SNLS 03D4cj}, \object{SNLS 03D4ag}, 
\object{SNLS 03D1cm}, \object{SNLS 04D3mk}, 
\object{SNLS 05D1by}, \object{SNLS 05D1hk}, \object{SNLS 05D3gy}. 
Supernova  SNLS~03D3bb is a super-Chandrasekhar Type Ia~\citep{bib:howell06}
with maximum light in the presurvey and 
does not pass the SN selection cuts. Supernova SNLS~03D1cm is a 1991T-like object, 
with $X_1 = 4.54$ and is therefore excluded by our cuts. Supernova SNLS~05D1by is 
rejected because of its red colour ($C=0.66$).
The other five events do not exhibit any sign of peculiarity in their light curves 
and are in our photometric sample. 

\subsection{Comparison with the SNLS 3-year cosmological sample}
In~\cite{bib:guy2010}, SNLS describes a set of well sampled light curves for 252
Type Ia supernovae that was used to derive the 3-year cosmological analyses of
the collaboration~\citep{bib:conley2010}. 
The photometric SN~Ia selection described in this paper recovers 172 events 
of that sample. The lost fraction (32\%) breaks down into 6\% due to SN 
selections, 16\% due to host galaxy redshift availability and 
10\% due to SN~Ia selections. 
Note that this result agrees with the efficiencies from synthetic SN~Ia light 
curves that we quoted in Section~\ref{sec:discus}.

\subsection{Comparison with the SNLS sub-luminous SN~Ia sample}
\cite{bib:santiago} provides a photometrically selected sample of 
18 sub-luminous SNe~Ia from SNLS data with $z<0.6$. All but two of 
these SNe are included in the sample analysed in this paper. 
Among those 16 events, 7 were kept by our photometric selection and 
9 were rejected because of insufficient temporal sampling (4), $\chi^2$ cuts (2) 
or extreme values of SALT2 colour or $X_1$ parameters (3). 
Despite the use of a light curve fitter trained on normal SNe~Ia only, our 
photometric selection appears to have some efficiency on sub-luminous SNe~Ia
as well.

\subsection{Contamination by core-collapse supernovae}\label{sec:conta}
Our set of 485 photometric SNe~Ia candidates includes  175 spectroscopically
identified SNe~Ia (see Table~\ref{tab:selIa}). 
One other event was spectroscopically identified 
as a possible core-collapse supernova initially, but was later found to be
a SN~Ia after it was discovered that the wrong spectrum had been extracted.
The remaining 309 events have no conclusive spectroscopic typing either because
no spectra were obtained or because of insufficient spectral signal-to-noise.
In this section we will estimate how many of the 485 events are, in fact
CC SNe that were incorrectly identified photometrically as SNe~Ia.
Core-collapse supernovae are typically more than a magnitude
fainter than SNe~Ia so a CC SN is unlikely to be confused with a SN~Ia
unless its redshift is significantly overestimated so that
its faintness is explained by its distance.

It is relatively simple to place an upper limit on the contamination
at low-redshift, $z<0.4$. Supernovae at these redshifts are spectroscopically
unambiguous because of the visibility of the $615\,{\rm nm}$ Si II absorption
feature characteristic of SNe~Ia.
While spectra for all SNLS CC~SNe were not obtained, those that appeared
to be possible SNe~Ia during the early phase of the explosion
were given high priority for spectroscopy.  We can therefore
expect that for $z<0.4$ we have a complete spectroscopic sample of those
CC~SNe most likely to be photometrically selected as SNe~Ia.
Our photometric sample at $z<0.4$ contains 35 spectroscopically identified SNe~Ia
and no spectroscopically identified CC SN, which gives a 95\% CL upper bound 
of 3 core-collapse supernovae. 
The contamination of our sample at low redshift is thus expected to be less than 
$\sim 3/35\sim 9\%$ at the 95\% CL.
This assumes that spectroscopic efficiencies at these redshifts are similar for Type Ia 
supernovae and for those among the CC events that would appear as possible SNe~Ia. 

To refine this estimate and extend it to higher redshifts,
we build a light curve model for CC~SNe based on the SNLS sample of low-redshift 
($z<0.4$) supernovae, using the analytical model of equation~(\ref{fitformula}) 
to describe the light curves.
To set up the simulation, we defined, as a CC test-sample, the set of 117 
low redshift events used in~\cite{bib:ratecc} to measure the CC rate
at $z<0.4$.  The 117 events include the 33 spectroscopically identified
CC~SNe and 84 events with no spectra and not photometrically selected as SNe~Ia.
To these, we added 9 spectroscopically identified 
CC~SNe that had either no host galaxy photometric redshift or 
a host galaxy photometric redshift above 0.4.

\begin{figure}[hbtp]
\begin{center}
\begin{tabular}{c}
\epsfig{figure=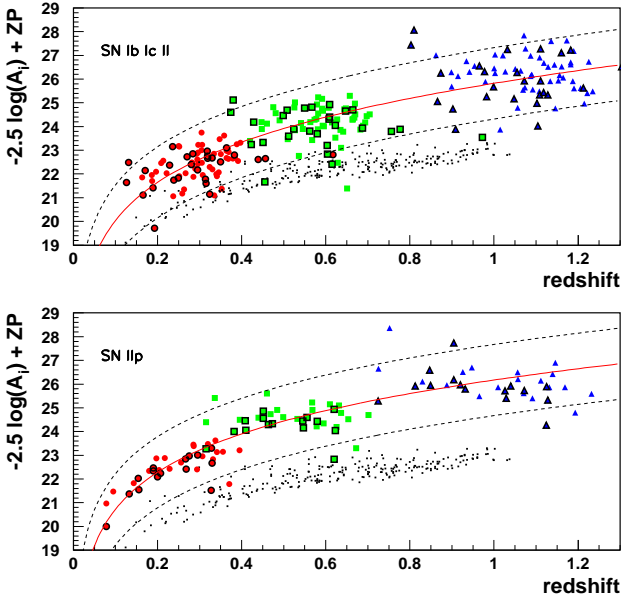,width=\columnwidth} 
\end{tabular}
\caption[]{Magnitude corresponding to $A_i$, the $i_M$-band 
amplitude parameter of the model representing SN light curves, as a function of 
redshift, for fast-declining CC supernovae (top) and plateau ones 
(bottom), as measured in the 3-year SNLS data.
Red dots come from fits to $i_M$-band light curves, 
green squares (resp. blue triangles) from fits in the $r_M$ (resp. $g_M$) filter
taken as estimates of $A_i$ at higher redshifts (see text). 
Symbols with black borders indicate spectroscopically identified events.
The curves are fits to the average redshift dependence of the magnitude, 
excluding points outside the dashed lines.
As a comparison, black dots give magnitudes from spectroscopically 
identified SNe~Ia.
}
\label{fig:simccamp}\end{center} 
\end{figure}

As detailed in section~\ref{sec:presel}, the analytical model of 
equation~(\ref{fitformula}) describes the signal with four parameters 
in each filter, namely an amplitude, a date of maximum, a fall time and 
a rise time.
Distributions of these parameters in the $i_M$ filter, as well as
correlations between these and the parameters in each of the other filters 
were measured in the test-sample and parametrised for their subsequent 
use in the light curve simulation. This was done separately for events
with plateau light curves (35 events, either events spectroscopically confirmed as
SNIIP or events with fall times above 60 days) and for fast-declining
events (77 events, either events spectroscopically identified as SN~Ib, SN~Ic or SN~II
or events with fall times below 40 days). 
No further distinction could be made in the latter category between Type Ib/c and 
Type II supernova light curves.

The test-sample is mostly limited at low redshift, below $z~\sim~0.4$.
In order to model the average redshift dependence of the amplitudes, 
the parameters of the low-redshift supernovae were redshifted to mimic
supernovae at higher redshifts.
The amplitudes in the $r_M$ and $g_M$ bands at a low redshift $z$ 
were thus taken as estimates of $A_i$ at a higher redshift $z_i$ defined as:
\[ 1 + z_i \equiv \frac{\lambda_i}{\lambda} (1+z) \]
after rescaling by:
\[ A_i \equiv A \frac{\lambda_i}{\lambda} \frac{d_L(z)^2}{d_L(z_i)^2} \]
where $\lambda_i$ and $\lambda$ are the mean wavelengths of the $i_M$ and  
$r_M$ or $g_M$ filters and $d_L(z)$ is the luminosity distance for the 
cosmology introduced in Section~\ref{sec:lcsimul}. 
In the above formulas, $z$ was taken as the spectroscopic redshift whenever 
available (50\% of the cases), and as the host galaxy photometric redshift 
otherwise. 
Amplitudes in the $r_M$ and $g_M$ filters were thus used to estimate 
the $i_M$ band amplitude  $A_i$
at redshifts $0.4<z_i<0.7$ and $0.7<z_i<1.2$, respectively.
Figure~\ref{fig:simccamp} presents the estimated $A_i$ 
as a function of $z_i$. 

The average redshift dependence of the modelled $i_M$-band magnitude was fitted 
with the logarithm of a third order polynomial in the redshift $z_i$. 
The test-sample of plateau (resp. other CC) supernovae has a scatter of 0.5~mag
(resp. 0.8~mag) around this average behaviour.  
Given the above scatters in magnitude, the dependence of our $A_i$ modelling 
with the reference cosmology assumed in the luminosity distance computation was 
found to be negligible.
The fitted redshift dependence of the modelled $i_M$-band magnitude 
and the spread of the residuals around the latter were used 
to simulate $i_M$-band amplitudes according to a Gaussian distribution.
Amplitudes in the other bands were simulated to reproduce the observed colours in
the test-sample. The colours simulated at high redshift were checked to be identical
to those for lower redshift supernovae in the appropriate (bluer) bands.

\begin{figure}[htbp]
\begin{center}
\epsfig{figure=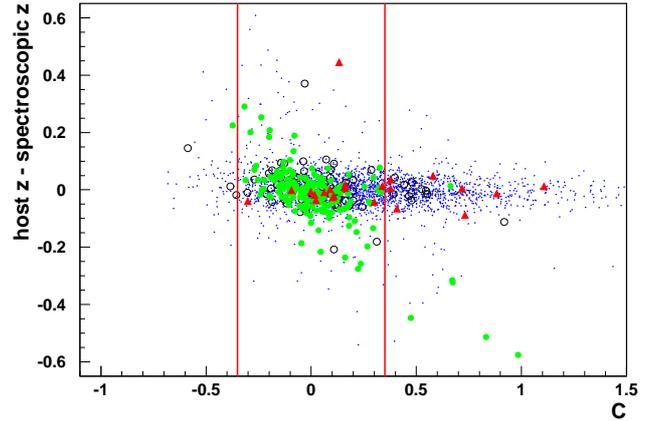,width=\columnwidth} 
\caption[]{Difference between redshift assignments as a function of the SALT2 
fitted colour, for synthetic core-collapse events (blue dots) and 
data events (all other symbols), after all selections previous to that 
on SALT2 $X_1$ and colour (see text). 
Green circles (resp. red triangles) are data events present in our
sample which have been identified as Type Ia (resp. core-collapse) 
supernovae by spectroscopy. Open black circles stand for data events 
with no spectroscopic identification.}
\label{fig:simccdz}\end{center} 
\end{figure}

\begin{figure}[htbp]
\begin{center}
\epsfig{figure=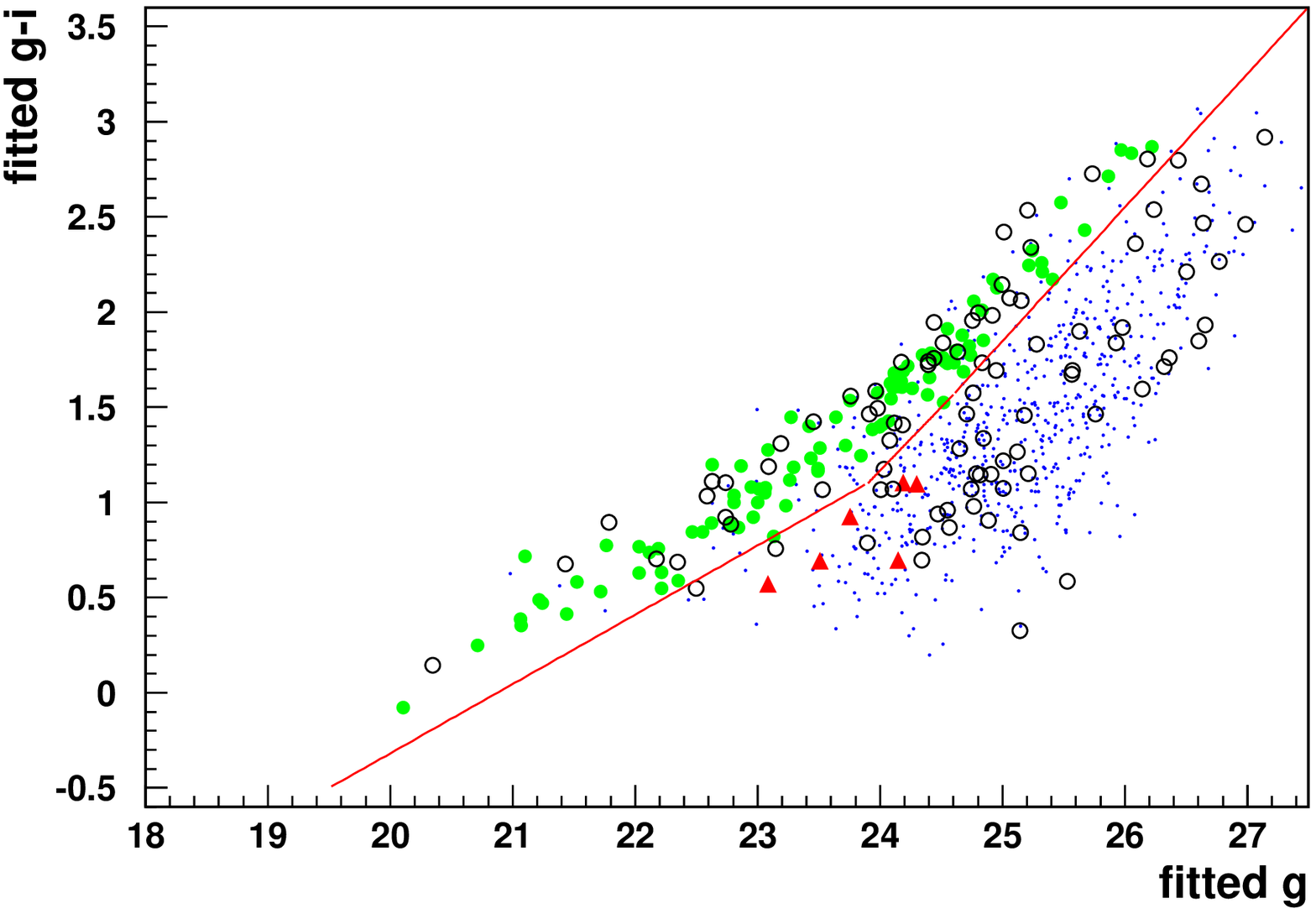,width=\columnwidth} 
\vskip -0.9cm
\epsfig{figure=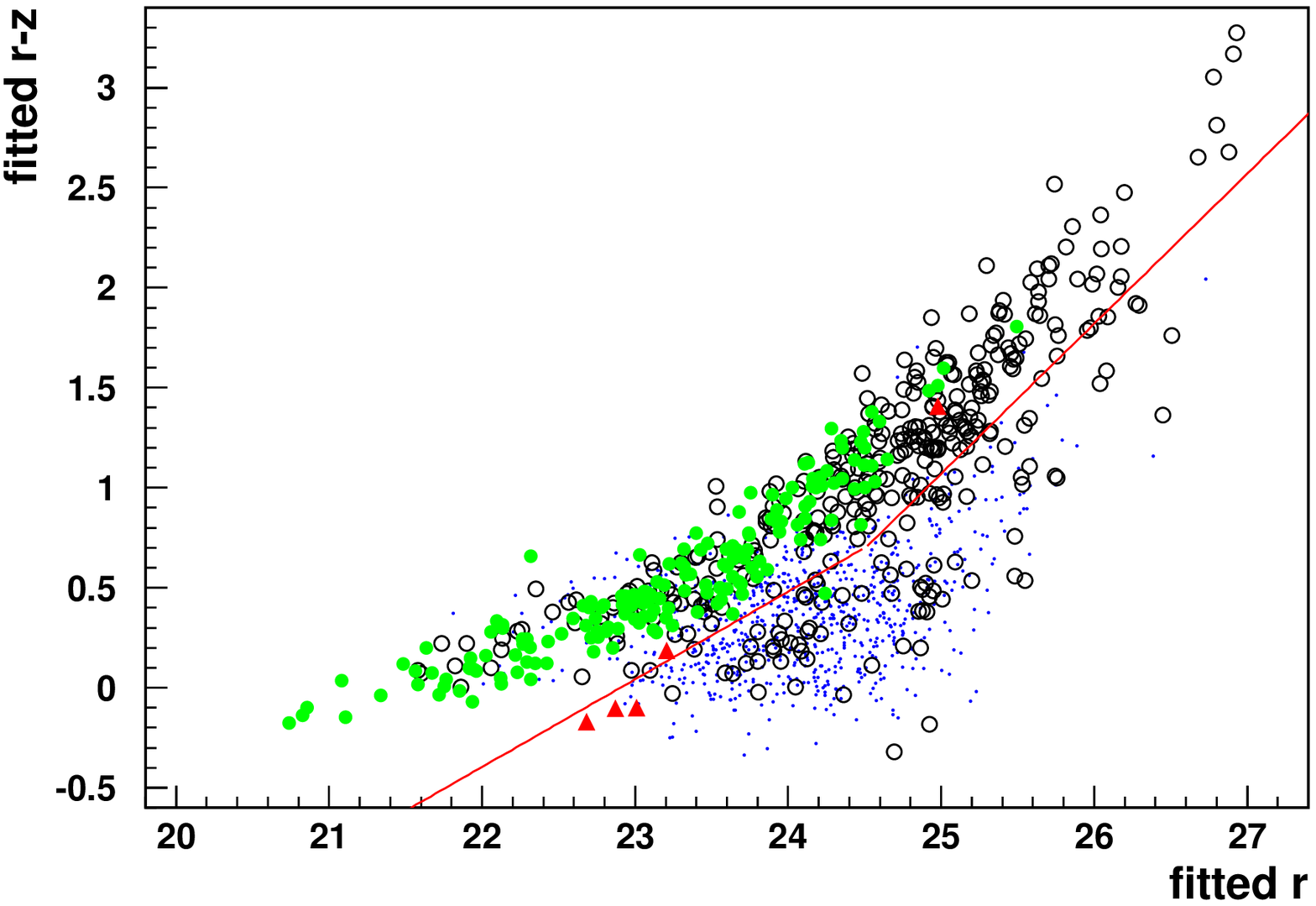,width=\columnwidth}  
\vskip -0.9cm
\epsfig{figure=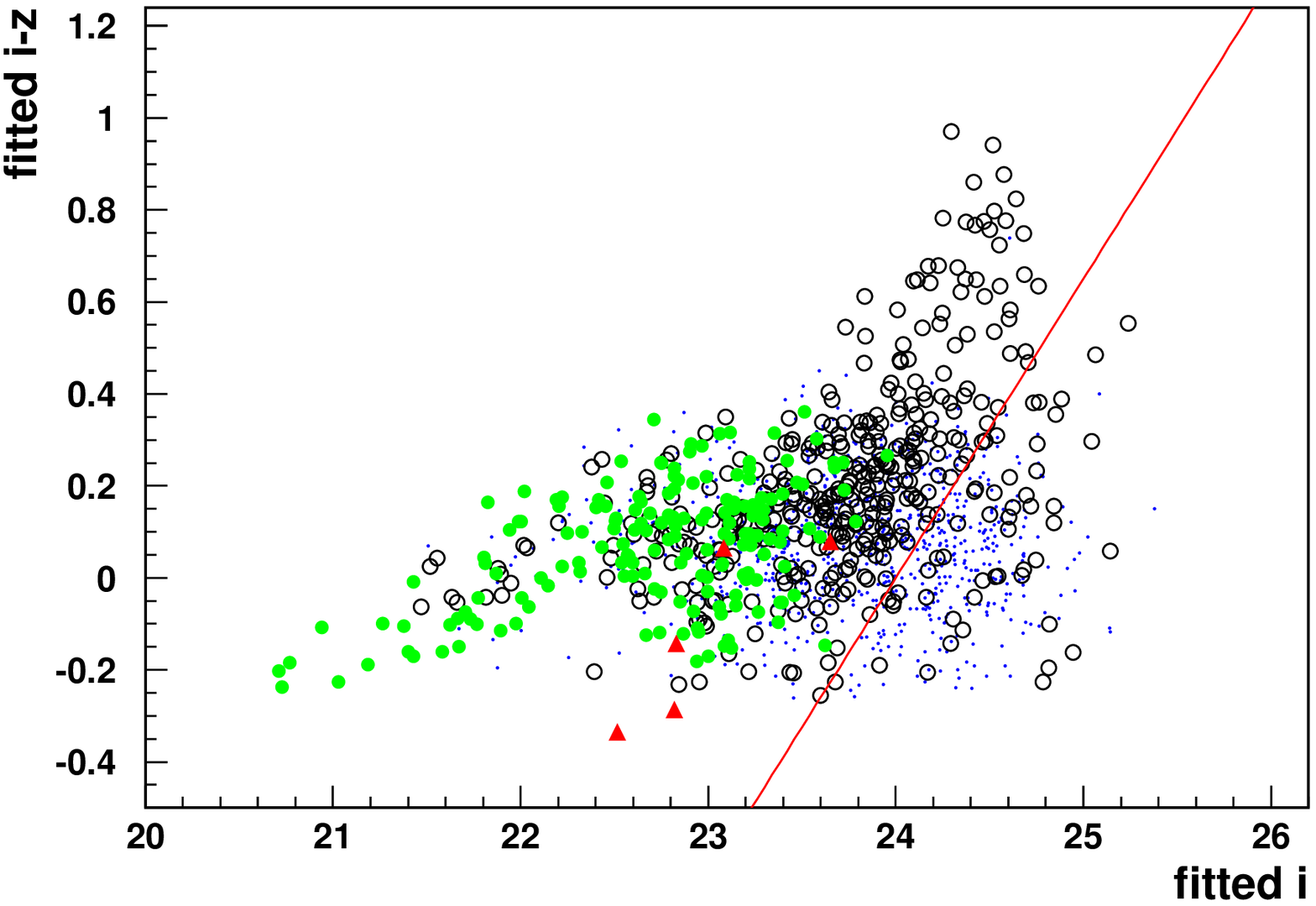,width=\columnwidth}
\caption[]{Colour-magnitude diagrams from SALT2 fitted magnitudes used
in the photometric SN~Ia selection, for synthetic core-collapse events and 
data events that have passed all selections prior to those based on these diagrams.
Same colour code as in Figure~\ref{fig:simccdz}.}
\label{fig:simccmag}\end{center} 
\end{figure}

We generated synthetic core-collapse light curves in the redshift range $0<z<1.2$.
As in the SN~Ia simulation,
host galaxy photometric redshifts were also generated, in order to reproduce
the selections of the analysis. 
A total of 40,000 light curves was simulated (20,000 in each category), 
of which 13\% remained at the end of the SN selection.
To illustrate the output of the simulation,
Figures~\ref{fig:simccdz} and~\ref{fig:simccmag} show
some of the constraints used in the SN~Ia selection,
with SNLS data now compared with synthetic CC light curves. 
We observe that our simulation based on low-redshift events
reproduces the distribution of spectroscopically identified CC~SNe
and of supernovae with no spectra.

The contamination of the photometric sample of SN~Ia candidates by 
CC~SNe was then derived as follows.
Our synthetic light curve simulation assumed redshift-independent volumetric
supernova distributions. To account for this,
synthetic events were first weighted by the actual redshift dependence expected
for the volumetric explosion rates,
assuming the SN~Ia rate to be proportional to $(1+z)^2$~\citep{bib:pritchet} and 
the core-collapse SN rates to vary as the star formation rate, i.e.
as $(1+z)^{3.6}$~\citep{bib:hopkins}. 
The numbers of weighted events passing the photometric SN~Ia selections were then 
computed for the three samples of synthetic light curves (one for SNe~Ia, one for 
SNe~IIP and one for all other core-collapse SNe) in the redshift range $z<1.2$.
To normalise the SN~CC and Ia simulations with each other, we used the
CC to Ia volumetric rate ratio published in~\cite{bib:ratecc}, that holds for
low redshift ($z<0.4$) events, in an absolute magnitude interval extending to 
4.5 mags fainter than normal SNe~Ia. The normalisation was set in order that
weighted synthetic events reproduce the published ratio in the same conditions 
of redshift and magnitude.

After normalisation, we obtained a contamination of $17.6 \pm 3.9(stat.)$ 
supernovae in our photometric sample of 485 SN~Ia candidates. 
Essentially all (99\%) of the contaminating events are non-plateau
core-collapse supernovae.
The statistical  uncertainty is mostly due to the small number of
events in the sample of low-redshift CC~SNe which led to the 
measurement of the CC to Ia volumetric rate ratio used in the above 
normalisation. The statistical uncertainty on that measurement was
20\%.

\begin{table}[htb]
\caption[]{Contamination from core-collapse supernovae.}
\label{tab:selCC}
\begin{center}
\begin{tabular}{ccccc} \hline \hline
Redshift & Events & CC \\ \hline
$z_{gal}<0.4$      &  49 & 4.4$\pm$ 1.1   \\
$0.4<z_{gal}<0.8$  & 196 & 11.1$\pm$ 2.5   \\
$0.8<z_{gal}$      & 240 & 2.1$\pm$ 0.5  \\
\hline
\end{tabular}
\tablefoot{The table shows the
number of events passing the photometric SN~Ia selection and the estimated
CC contamination in three bins of host galaxy 
photometric redshift. } 
\end{center}
\end{table}

Note that the synthetic CC events remaining at the end of the selections 
are at moderate host galaxy photometric redshift, $z_{gal}\sim0.5$ on average
with an r.m.s. of 0.2 (see also Table~\ref{tab:selCC}) 
and, when reconstructed as SNe Ia by SALT2, 
exhibit positive colours (with a mean value of 0.15 and an r.m.s. of the 
same order) and a broad distribution of $X_1$ values 
(with a mean of -0.3 and an r.m.s. of 1.7). 
Moreover, the rate of 5$\sigma$ outliers in host galaxy photometric redshift is 5\% in this
sample, slightly higher than the 1.4\% observed in the whole photometric sample.

\section{Comparison of event subsamples with and without spectroscopic 
identification}\label{sec:char} 

\begin{figure}[htbp]
\begin{center}
\begin{tabular}{c}
\epsfig{figure=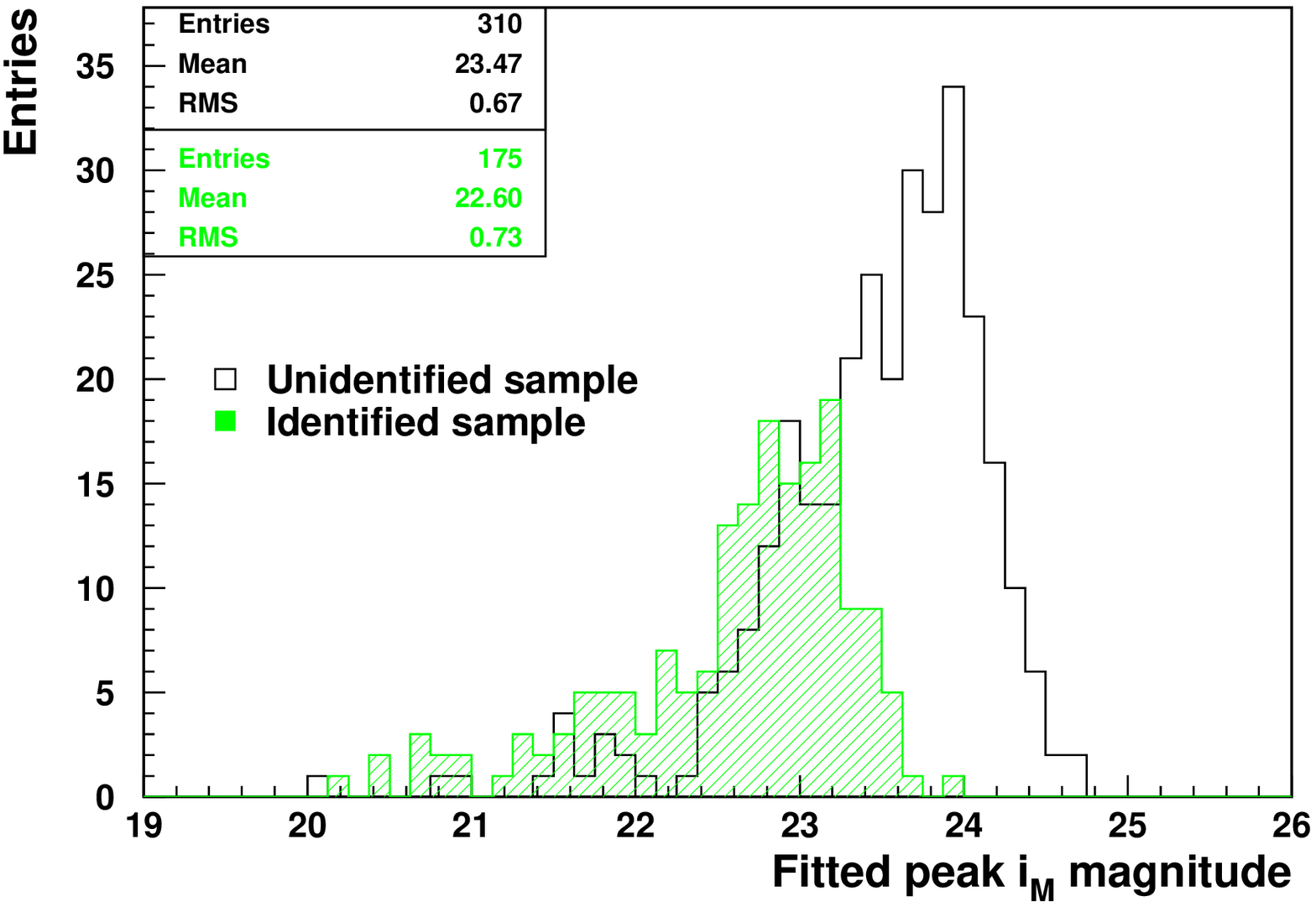,width=\columnwidth} \\
\epsfig{figure=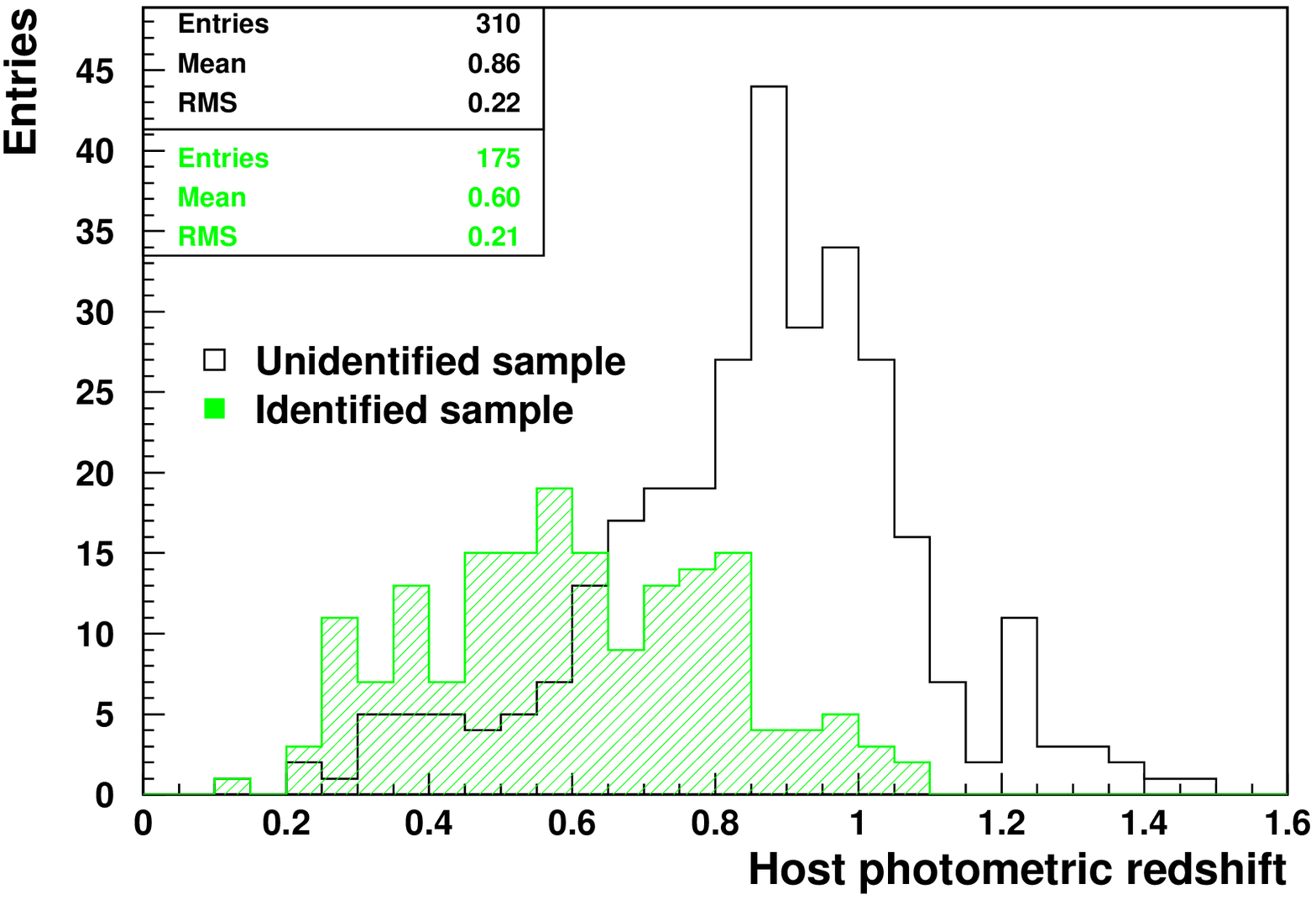,width=\columnwidth}
\end{tabular}
\caption[]{Distributions of the SALT2 fitted $i_M$ peak magnitude (top) 
and of the host galaxy photometric redshift (bottom) for the identified (in green) 
and unidentified (in black) subsamples of photometrically selected SNe~Ia.}
\label{fig:imag}\end{center} 
\end{figure}

The sample of 485 SN~Ia photometric candidates contains 175 spectroscopically
identified SNe~Ia, called hereafter the ``identified sample'',
and 310 events with no spectral confirmation\footnote{Note that 
event SNLS~06D2bo has been included in the unidentified sample.}, 
called hereafter the ``unidentified sample''.
The distribution of the peak magnitude in $i_M$ for the two subsamples is given in 
Figure~\ref{fig:imag}. The unidentified sample is about one magnitude deeper on average
than the identified sample, 
as expected from the constraints to obtain a spectrum for the identified SNe~Ia. 
This translates into an average redshift of 0.60 for the 
identified set and of 0.86 for the additional set. 

\subsection{Comparison of bright events}
At bright magnitudes, $i_M<23$, 
there are 110 identified events and 56 unidentified ones, all but 5 events found in
the real time analysis. 
We traced back the reasons why these bright events missed a spectroscopic 
identification.
A third of them had been detected in the real-time analysis, 
declared (from photometry) as probable SNe~Ia and sent for
spectroscopy. Insufficient quality of the spectrum, however, resulted in a spectroscopic 
redshift (usually from the host galaxy) but no typing of the supernova. 
Another third  were also declared as probable SNe~Ia but could not be scheduled 
for spectroscopic follow-up near maximum light.
The last third were usually less convincing candidates according to their real-time,
and thus partial, light curves. 

\begin{table}[htb]
\caption[]{Comparison of events with and without spectroscopic identification.}
\label{tab:compareparams}
\begin{center}
\begin{tabular}{cccc} \hline \hline
Parameter & Identified  & Unidentified  & KS \\
          & sample & sample &  proba \\ \hline
$C$       & $0.00\pm0.01$ & $0.02\pm 0.02$  & 0.40 \\
$X_1$     &   $0.23\pm0.08$   &  $-0.01\pm 0.13$  & 0.30 \\
\hline
\end{tabular}
\tablefoot{
The mean values of SALT2 colour and $X_1$ are given for photometrically 
selected events with $i_M<23$, split into subsamples with and without spectroscopic
identification. 
Numbers in the last column are the Kolmogorov-Smirnov probabilities that the two 
distributions arise from the same parent distribution.
} 
\end{center}
\end{table}
 
\begin{figure}[htbp]
\begin{center}
\begin{tabular}{c}
\epsfig{figure=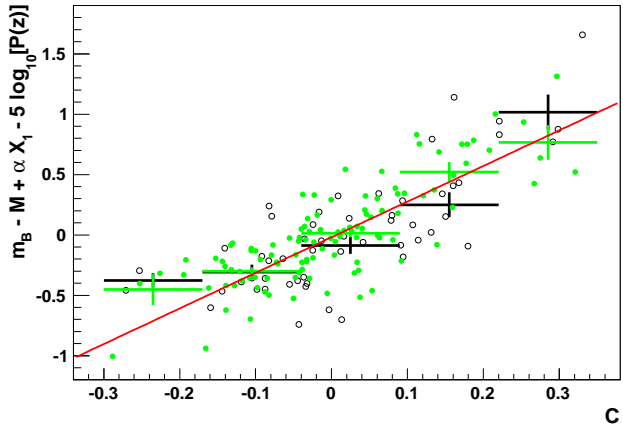,width=\columnwidth} \\
\epsfig{figure=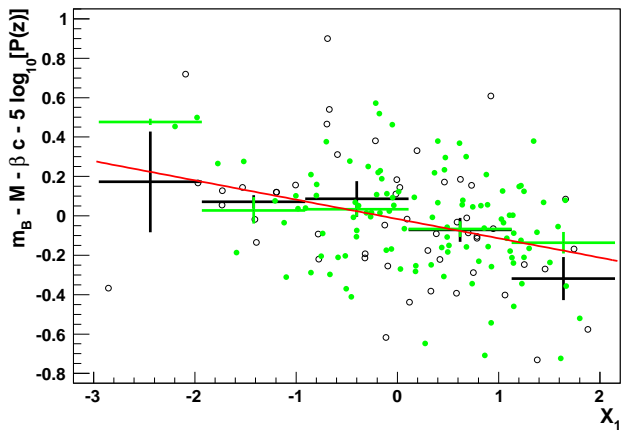,width=\columnwidth} 
\end{tabular}
\caption[]{Partial Hubble diagram residuals (and profiles) as a function of colour (top)
and $X_1$ (bottom) for the identified (green filled circles and profiles) and 
unidentified (black open circles and profiles) subsamples of photometrically 
selected SNe~Ia with $i_M<23$.
The straight lines are linear fits to the full sample of events with $i_M<23$.}
\label{fig:brighterbluer}
\end{center} 
\end{figure}

The characteristics of the two subsamples of bright events were compared,
based on their complete light curves as reconstructed in this analysis.
Their colour and $X_1$ distributions were found
to be compatible, as summarised in Table~\ref{tab:compareparams}. 
We also compared the colour-magnitude and $X_1$-magnitude relations in the two
subsamples.
For this purpose, the distance modulus of each event was defined from the SALT2 
fitted $B$-band peak magnitude $m_B^*$, $X_1$ and colour $C$ as:
\begin{equation}
\mu_B = m_B^*-M+\alpha X_1 - \beta C\;\;
\label{modulus}
\end{equation}
using values of $M$, $\alpha$ and $\beta$ introduced in Section~\ref{sec:lcsimul}.
Residuals were then computed from these distance moduli by subtracting 
$5 \log [P(z)]$ where $P(z)$ is a third-degree polynomial that we fitted
on the total photometric sample in order to describe
the mean redshift dependence of the distance modulus 
with no assumption on a specific cosmology model.
The Hubble diagram residuals without the colour or $X_1$ term in the distance 
modulus are represented in Fig.~\ref{fig:brighterbluer}.
The figure shows the fit of the colour-magnitude and $X_1$-magnitude relations
from the full sample of bright events, as fits from the two subsamples were found
to be indistinguishable. 

\par 
In conclusion, when restricted to the same range of $i_M$ peak magnitudes,
the unidentified and identified samples of events
do not exhibit significant differences.

\subsection{Comparison of full samples}\label{sec:bias}

We now consider the full set of photometrically selected SNe~Ia 
as an extension of the identified subsample towards fainter events,
and estimate the impact on distance moduli of using the limited
sample of spectroscopically identified SNe~Ia. 
The Malmquist bias due to spectroscopic sample selections is an
important issue in cosmology fits and was thoroughly studied in
SNLS with Monte Carlo simulations. 
The results, reported in~\cite{bib:perrett} were used to correct SN~Ia 
distance moduli in SNLS 3-year cosmological 
analyses~\citep{bib:guy2010,bib:conley2010}.
The aim here is to check whether we can measure this bias directly from data 
and with an analysis completely independent from that used to define the 
3-year SNLS cosmological sample.

To do so, ideally one would compare average distance moduli at a 
given redshift, measured from the identified subsample and from the whole 
photometric sample.
Our statistical sample being limited, we compare Hubble diagram residuals
in large bins of redshifts instead of distances at given redshift values.
As all events were submitted to the same selection procedure and the same 
redshift assignment, such a comparison is 
directly sensitive to different Malmquist biases and should not be altered by 
systematic uncertainties related to data processing. 

\begin{figure}[hbtp]
\begin{center}
\epsfig{figure=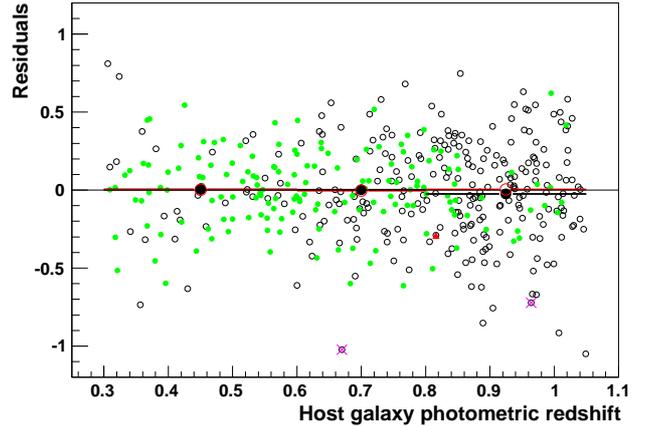,width=\columnwidth}
\caption[]{Hubble diagram residuals.
Green dots stand for the identified SN~Ia subsample and open black circles 
for the unidentified subsample. Means are shown in three redshift bins, 
black for identified events, red for full sample. 
The red upward triangle is event SNLS~06D2bo which was erroneously typed
as a possible core-collapse supernova initially (see Section~\ref{sec:compa_rta}
for details).
The pink crosses mark the two $3\sigma$ outliers, which were excluded 
in the computation of the means.}
\label{fig:hubble}\end{center} 
\end{figure}

The residuals of the Hubble diagram are presented in Fig.~\ref{fig:hubble}
for the two samples.
The residuals were again defined from the distance moduli in Eq.~\ref{modulus} 
by subtracting $5 \log [P(z)]$ where $P(z)$ is the third-order polynomial 
introduced in the previous section.
The Hubble diagram residuals were then averaged in three large redshift bins, $[0.3-0.6]$, 
$[0.6-0.8]$ and $[0.8-1.05]$, where both identified and unidentified samples 
had enough events to allow a quantitative comparison. In the computation of the
means, 3$\sigma$ outliers were rejected.

\begin{table*}[hbtp]
\caption{Hubble diagram mean residuals for different samples of SN~Ia data selected by this analysis.}
\label{tab:resi}
\begin{center}
\begin{tabular}{cccc} \hline \hline
SN Ia sample & \multicolumn{3}{c}{Mean residuals in redshift bins} \\ 
             & $0.3<z<0.6$     & $0.6<z<0.8$ & $0.8<z<1.05$  \\ \hline
Identified   & $0.004 \pm 0.032$ (76) & $0.003\pm 0.033$ (51)& $-0.025\pm 0.041$ (31) \\
Unidentified & $-0.002\pm0.053$ (31)  & $-0.009\pm 0.032$ (67)& $0.013\pm 0.022$ (160) \\
Total        &$0.002\pm 0.027$ (107) & $-0.003 \pm 0.023$ (118) & $0.005\pm 0.019$ (191)\\
\hline
$\Delta \mu_{id}$ = Identified - Total & $0.002\pm 0.017$ & $0.006\pm 0.024$ & $-0.030\pm 0.036$  \\
\hline
\end{tabular}
\tablefoot{
Host photometric redshifts were used for all events. 
The number of events in each bin is indicated in parenthesis. 
The last line shows the difference $\Delta \mu$ between the mean residuals 
in the identified and total samples. Uncertainties are statistical only.}
\end{center}
\end{table*}

\par  
Mean values of the residuals (each weighted by the inverse of its variance) 
are reported in Table~\ref{tab:resi}. The variance on each point includes both a 
contribution from the light curve fit and one from the use of a photometric redshift. 
For the latter, we used a typical redshift uncertainty of $\delta_z = 0.05$, that we
translated into a distance modulus uncertainty using the derivative 
of the cosmological model introduced in Section~\ref{sec:lcsimul}.
By construction, the residuals of the total sample are 0 (see also 
Figure~\ref{fig:hubble}).
The last line of Table~\ref{tab:resi} gives the difference between the mean residuals 
in the identified and total samples, which directly measures the mean
difference in distance modulus between the two samples, $\Delta \mu_{id}$.

\par 
Up to a redshift of 0.8, distance moduli in the identified and the 
total samples do not exhibit significant differences.
In the last redshift bin, corresponding to a mean redshift of 0.9, 
distance moduli in the two samples have an average offset of 
$\Delta\mu_{id} =-0.030\pm 0.036$~ (stat.).  This result is stable 
with respect to different redshift binning or outlier cuts. 

\subsection{Offset correction and systematics}

\par 
\begin{figure}[htbp]
\begin{center}
\begin{tabular}{c}
\epsfig{figure=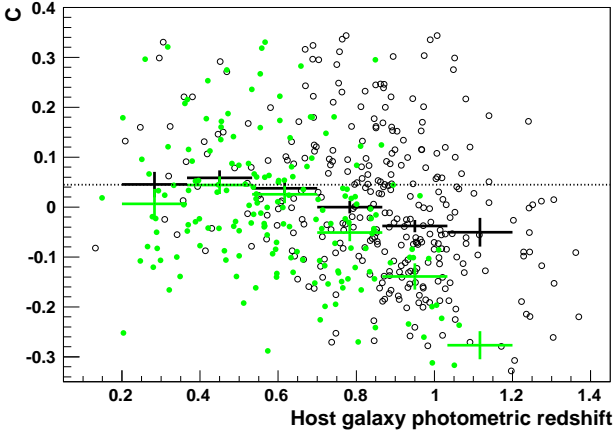,width=\columnwidth} \\
\epsfig{figure=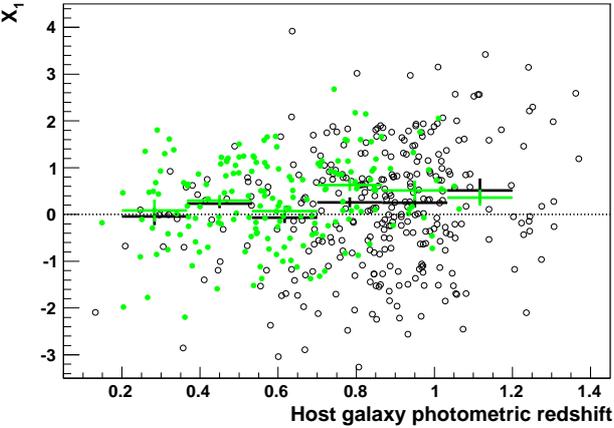,width=\columnwidth}
\end{tabular}
\caption[]{Distributions of colour (top) and $X_1$ (bottom) as a function
of the host galaxy photometric redshift,
for the identified (green filled circles and profiles) and unidentified (black open 
circles and profiles) subsamples of photometrically selected SNe~Ia. 
}
\label{fig:charallmag}\end{center} 
\end{figure}

As shown in Figure~\ref{fig:charallmag}, the bias towards selecting
bluer, and hence brighter events at high redshift is clear in the
identified sample and attenuated in the photometric sample. Although 
the latter effect may be due to evolution of Type Ia supernovae,
it can also result from selection effects
(our photometric sample being magnitude limited and biased  
against SNe in faint host galaxies).
In that case, the offset value quoted in the previous section should 
be corrected for the incompleteness of the photometric sample. 
A correction for the residual core-collapse contamination of the sample 
may also be needed.

These two corrections were determined from the synthetic supernova light curves,
which were submitted to the same selections as data. 
Residuals were here computed w.r.t. the luminosity distance for 
the cosmology used in the simulation (see Section~\ref{sec:lcsimul}). 
In the redshift bin [0.80, 1.05], synthetic SNe~Ia passing our photometric
selection exhibit a negative average residual of $-0.0148 \pm 0.006 (stat.)$, 
while the selected CC~SNe (less than 1 event out of 191) contribute for a 
tiny residual of $+0.0005~\pm~0.0008 (stat.)$.
The offset of the photometric sample thus amounts to
$\Delta\mu_{phot}=-0.014~\pm~0.006 (stat.)$ 
and the total offset of the identified subsample in the redshift interval 
$0.8<z<1.05$ is 
$\Delta\mu=\Delta\mu_{id}+\Delta\mu_{phot}=-0.044\pm 0.036 (stat.)$.

\begin{table}[hbtp]
\caption{Effect of redshift and photometry resolutions on mean residuals.}
\label{tab:sys} 
\begin{center}
\begin{tabular}{ccc} \hline \hline
Analysis               &  \multicolumn{2}{c}{Mean residual} \\ 
conditions             & Bright SNe  & Photometric SNe  \\ 
\hline
This analysis          & $-0.033 \pm 0.007$  & $-0.015 \pm 0.006$  \\
\hline
$|\Delta z|/(1+z)<10\%$ & $-0.034 \pm 0.007$  & $-0.016 \pm 0.006$  \\
$|\Delta z|/(1+z)<3\%$  & $-0.042 \pm 0.009$  & $-0.020 \pm 0.008$  \\
$|\Delta z|/(1+z)<1\%$  & $-0.044 \pm 0.014$  & $-0.021 \pm 0.012$  \\
Perfect photom. \& z    & $-0.043 \pm 0.007$  & $-0.024 \pm 0.006$ \\
\hline
Difference              & $-0.010 \pm 0.009$   & $-0.009 \pm 0.008$ \\
\end{tabular}
\tablefoot{The table details the 
changes in the weighted mean residuals of the identified and photometric SN~Ia samples, 
as expected if resolutions in redshift and photometry were gradually 
improved w.r.t. their values in the present analysis (first line). 
All changes were estimated from synthetic SN~Ia light curves.
Uncertainties are statistical.}
\end{center}
\end{table}

\par Systematic uncertainties due to the limited precision of 
redshift assignment and photometry may affect this result. 
These effects were again studied with the synthetic SN~Ia light curves.
Changes in the mean residuals of the photometric sample and its bright 
subsample are reported in Table~\ref{tab:sys}. 
The difference between the residuals of the two samples compares 
directly with $\Delta\mu_{id}$, while the mean residual of the photometric
sample identifies with the correction term $\Delta\mu_{phot}$.
Table~\ref{tab:sys} shows that improving the redshift determination
and the photometry precision leads to almost identical shifts of 
the mean residuals of the two samples. These two effects have therefore a 
negligible impact on the measurement of $\Delta\mu_{id}$. 
The table also shows that
the systematic uncertainty on $\Delta\mu_{phot}$ is of order -0.01,
negligible w.r.t. the statistical uncertainty of the method.

As previously mentioned, 
a precise determination of the magnitude offset in the SNLS 3-year 
spectroscopic sample was obtained in~\cite{bib:perrett} with 
Monte Carlo simulations.
The shift in the average magnitude of the spectroscopic sample towards 
brighter values was found to become significant for $z>0.75$, rising 
from $-0.013\pm0.001 (stat.)$ at $z\sim0.8$ to 
$-0.038\pm0.003 (stat.)$ at $z\sim1.05$, with a systematic uncertainty 
of 20\%.
The result reported here, 
$\Delta\mu=-0.044\pm 0.036 (stat.) \pm 0.010 (syst.)$ in the redshift interval 
$0.8<z<1.05$,
obtained mostly from data and from
an analysis independent from the real-time SNLS processing, agrees
with the above values.

\section{Conclusions}
 
The SuperNova Legacy Survey offers one of the best data sets to test SN~Ia 
photometric 
typing methods in the redshift range between 0.2 and 1.0.
Although spectra remain essential to achieve the level of accuracy required in
cosmology studies, many supernova studies such as rate measurements or 
correlations with host galaxy properties benefit from the larger statistics
allowed by photometrically identified samples.

In this paper, the 3-year SNLS image sample was submitted to a deferred search 
for Type Ia supernovae based on their multi-colour light curves
and an external catalogue of photometric redshifts of host galaxies. 
SN-like transient events were first selected using light curve shape criteria.
The selected light curves were fitted under the
assumption that the events were SNe~Ia using SALT2 as light curve fitter. 
Selections were then designed to both discriminate SNe~Ia against core-collapse 
ones 
and to assess the reliability of the redshift assignment. They were
applied to the SALT2 fitted parameters, rest-frame $B-V$ colour, 
stretch-related $X_1$ parameter, as well as colours and magnitudes in the 
MegaCam filters. These selections allowed us
to select a sample of 485 SN~Ia events with host galaxy photometric redshifts.

The performance of the analysis was studied with synthetic light curves of
Type Ia and core-collapse supernovae. The selection efficiency for bright 
SNe~Ia with well sampled light curves is 80\% in case a host galaxy
photometric redshift is available. The contamination of the selected sample
from other types of supernovae is 4\%.

One third of the photometrically selected events have 
confirmed Ia types from SNLS spectroscopy. These events were
compared with the rest of the photometric sample which has no or incomplete
spectroscopic information.
In the range of magnitudes allowed by spectroscopic
identification, there is no significant differences in the  
intrinsic features of events with and without spectroscopic 
identification. 

The higher magnitude limit of the whole photometric sample allowed distances 
derived from spectroscopically identified events to be tested in order to
measure the Malmquist bias due to spectroscopic sample selections,
directly from data and with an analysis independent from the
SNLS cosmological analyses.
A magnitude offset was found at redshifts above 0.8,
though with a large statistical uncertainty. 
The obtained offset is consistent 
with the more precise result derived from Monte Carlo simulations 
which was used to correct SN~Ia distance moduli in the SNLS 3-year 
cosmological analyses.

This paper demonstrates the feasibility of a photometric selection of high 
redshift supernovae with known photometric host galaxy redshifts, opening 
interesting prospects for cosmological analyses from large photometric 
SN~Ia surveys, as planned in the near future. 
In this respect, the points we found most important for such an 
analysis are the following. 
The cadence of observations must be chosen to obtain light curves 
with a temporal sampling of a few days. 
The number of filters must allow SN colours and magnitudes to be 
determined in a reliable way on the whole SN redshift range.
A low contamination by core-collapse supernovae can be achieved when using
photometric host galaxy redshifts with a few \% resolution and rate
of catastrophic redshifts, provided selections include enough criteria
to reject events with incorrect redshift assignments.

\subsection*{Acknowledgements} 

SNLS is based on observations obtained with MegaPrime/MegaCam, 
a joint project of CFHT 
and CEA/Irfu, at the Canada-France-Hawaii Telescope (CFHT) which is operated 
by the National Research Council (NRC) of Canada, the Institut National des 
Sciences de l'Univers of the Centre National de la Recherche 
Scientifique (CNRS) 
of France, and the University of Hawaii. This work is based in part on data 
products produced at TERAPIX and the Canadian Astronomy Data Centre as part 
of the Canada-France-Hawaii Telescope Legacy Survey, a collaborative project 
of NRC and CNRS. 
This paper makes use of photometric redshifts produced jointly by TERAPIX and 
VVDS teams.
SNLS also relies on observations obtained at the European
Southern Observatory using the Very Large Telescope on
the Cerro Paranal (ESO Large Programme 171.A-0486), 
on observations (programs GN-2006B-Q-10, GN-2006A-Q-7, GN-2005B-Q-7, GS-2005B-Q-6, GN-2005
A-Q-11, GS-2005A-Q-11, GN-2004B-Q-16, GS-2004B-Q-31, GN-2004A-Q-19, GS-2004A-Q-11, GN-2003
B-Q-9, and GS-2003B-Q-8) obtained at the
Gemini Observatory, which is operated by the Association
of Universities for Research in Astronomy, Inc., under a
cooperative agreement with the NSF on behalf of the Gemini
partnership: the National Science Foundation (United States),
the Particle Physics and Astronomy Research Council (United
Kingdom), the National Research Council (Canada), CONICYT
(Chile), the Australian Research Council (Australia), CNPq
(Brazil) and CONICET (Argentina), and on observations
obtained at the W.M. Keck Observatory, which is operated
as a scientific partnership among the California Institute of
Technology, the University of California and the National
Aeronautics and Space Administration. The Observatory was
made possible by the generous financial support of the W.M.
Keck Foundation.
MS acknowledges support from the Royal Society.

\section*{Appendix A: detection efficiency model}

The detection efficiency for a supernova with date $t_{0i}$ of maximum 
light in $i_M$-band 
depends on its peak magnitude $m_{0i}$, on the seeing and sky background 
during nearby observing times $t_k$, and on the relative epochs $t_{0i}-t_k$. 
The dependence of the efficiency on these variables was studied with 
Monte Carlo images on D1. 

\begin{figure}[htbp]
\begin{center}
\epsfig{figure=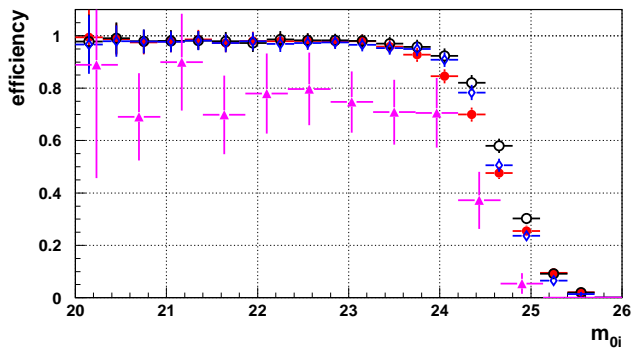,width = \columnwidth} 
\caption[]{Detection efficiency $\epsilon(m_{0i})$ on D1 derived from 
Monte-Carlo images as a function of the generated magnitude $m_{0i}$ 
at maximum light. 
Blue diamonds illustrate the efficiency for the first season of data 
(averaged over the full mosaic excluding CCD 3), red 
disks are for the second season (full mosaic), open black circles are for the 
third season (full mosaic), and pink triangles are for the 
supernovae occurring during the first season on CCD 3.} 
\label{fig:effMC}
\end{center}
\end{figure}

As an example, the season-averaged average detection efficiency 
$\epsilon(m_{0i})$ is illustrated in Figure~\ref{fig:effMC}.
On average over the CCD mosaic, all three seasons exhibit very similar 
efficiencies,
except for CCD 3 which failed to work during three out of the six dark 
time periods of the
first season. The efficiency is nearly magnitude-independent out to 
$m_{0i}=23.5$ and then steeply declines at fainter magnitudes.

\begin{figure}[htbp]
\begin{center}
\epsfig{figure=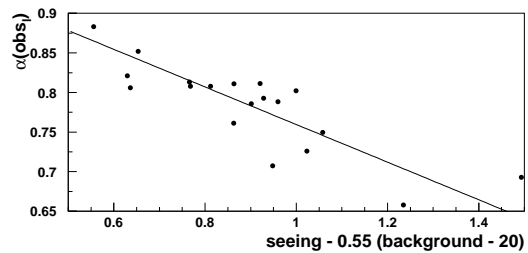,width = 7.5cm} 
\caption[]{Lunation detection probabilities $\alpha(\vec{o}_l)$ from the D1 
Monte Carlo  as a function of mean lunation seeing (in arcsec) and background 
(in mag per arcsec$^2$).
The line is the adopted $\alpha(\vec{o}_l)$. } 
\label{fig:plun}
\end{center}
\end{figure} 
 
\begin{figure}[htbp]
\begin{center}
\epsfig{figure=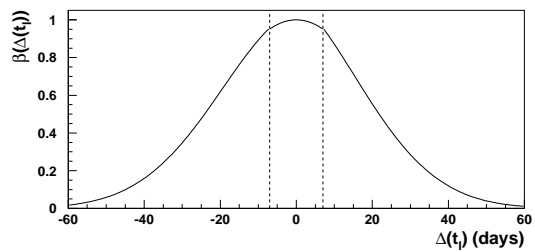,width = 7.5cm} 
\caption[]{Analytical model of the temporal dependence $\beta(\Delta t_l)$, 
here illustrated for a fourteen-day lunation. Dashed lines indicate the 
lunation boundaries. The model consists of two decreasing exponentials outside 
the lunation boundaries (with a steeper decline for positive $\Delta t_l$ 
because the fall time of a SN~Ia is larger than its rise time)
and is roughly flat inside.} 
\label{fig:beta}
\end{center}
\end{figure}

With varying observing conditions, we found that the efficiency as a 
function of magnitude
had the form of the mean efficiency in Figure~\ref{fig:effMC}, 
but that the maximum efficiency and cutoff 
magnitude depended on conditions. The adopted efficiency model therefore 
has the form:
\begin{displaymath}
\epsilon(t_{0i}) = \epsilon_{max}(t_{0i})\, F(m-m_c(t_{0i}))
\end{displaymath}
where $F(m-m_c)$ is the mean of the functions shown in Figure~\ref{fig:effMC}
($F(-m_c)=1$ and $F(0)=1/2$).  The function
$\epsilon_{max}$ is the  efficiency at bright magnitudes
while $m_c$ is the magnitude at which the efficiency drops to
$\epsilon_{max}/2$.
The  $t_{0i}$ dependence of $\epsilon_{max}$ and $m_c$
reflects the observing conditions near $t_{0i}$.
It was found that they could both be determined by a single 
function $f(t_{0i})$:

\begin{equation}
\epsilon_{\rm max}(t_{0i}) = 1 - 0.2 f(t_{0i}) \hspace{1cm}
m_c(t_{0i}) = 25.3 - 5.4 f(t_{0i})  \hspace{0.2cm} \label{eq:templ} 
\end{equation} 
The efficiency-loss function,  $f(t_{0i})$, vanishes for the best observing
conditions. It was  taken to be a product over all lunations:
\begin{eqnarray}
f(t_{0i})&=& \prod_l [1-\alpha(\vec{o}_l)\times\beta(\Delta t_l)] 
\label{eq:ft0}
\end{eqnarray}
where $\alpha$ is a function of the mean 
observing conditions, $\vec{o}_l=(seeing,\, sky\; background)$
for the lunation and $\beta$ is a function of the difference, $\Delta t_l$,
between the time of supernova maximum light and the
time of the nearest measurement.
Figures~\ref{fig:plun} and~\ref{fig:beta} show the adopted functions. 
Note that for $\Delta t_l > 50$~days, $\beta(\Delta t_l) \rightarrow 0$, so
in practice only the three lunations closest to the explosion determine  
$f(t_{0i})$.
The impact of temporary CCD failures is automatically accounted for in 
this model 
by setting $\alpha(\vec{o}_l)=0$ for non-working CCDs, leading to large 
intervals between 
the maximum light date $t_{0i}$ and the actual dates of observation, and 
hence to lower detection efficiencies.

\begin{figure}[htbp]
\begin{center}
\epsfig{figure=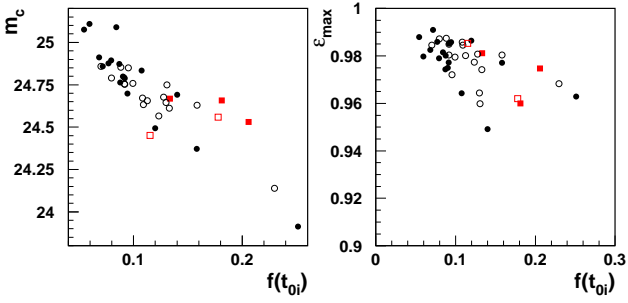,width = \columnwidth} 
\caption[]{Correlations between the measured parameters ($m_c$ and 
$\epsilon_{\rm max}$) and $f(t_{0i})$, as observed in the D1 Monte-Carlo.
Each point is an average over all $t_{0i}$ dates in a lunation (filled symbols)
or inter-lunation period (open symbols). 
Circles are averaged over the full mosaic, except for the first season
where CCD 3 has been singled out (red squares) due to intermittent failures.
Linear fits are used in the model to describe the observed correlations.
} 
\label{fig:emaxmcut}
\end{center}
\end{figure}

\begin{figure}[hbtp]
\begin{center}
\epsfig{figure=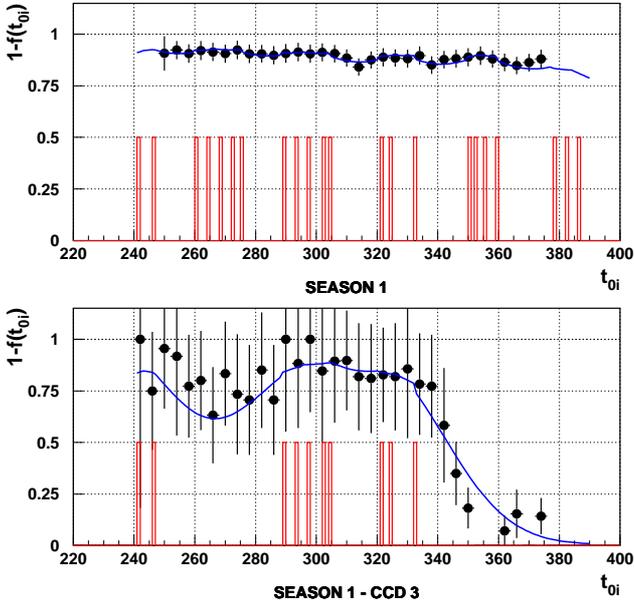,width =  \columnwidth} 
\caption[]{Temporal dependence of the detection efficiency, 
$1-f(t_{0i})$, on D1, with dates originating at January 0, 2003. 
The model (blue curve) is compared with Monte-Carlo estimates 
(black points). 
Actual observation dates are shown as the lower red 
histograms. The bottom plot is for CCD 3 which failed 
to work during the second and the last two lunations. The top plot is 
for the rest of the mosaic.} 
\label{fig:effD1_d}
\end{center}
\end{figure}

The linear dependence of $\epsilon_{\rm max}$ and $m_c$ on $f(t_{0i})$ 
(see equation~\ref{eq:templ}) is illustrated in Figure~\ref{fig:emaxmcut}.
We see that the primary effect of degraded observing conditions 
is to lower the cutoff magnitude:  conditions
that degrade $m_c\sim25$ to $m_c\sim24$ change 
the efficiency for bright supernova only from $\epsilon_{max}\sim0.98$
to $\epsilon_{max}\sim0.96$.
Note also that $m_c$ differs on average  by about 0.25~mag for
supernovae occurring  during a lunation compared to those
between lunations.
The case of CCD 3 in the first season has been also singled out to show
all cases of inter-lunation durations, from fifteen days
(normal CCD working conditions) to over one month.
Figure~\ref{fig:emaxmcut} shows that the observed correlations can be 
described by the same linear fits,
whether $t_{0i}$ occurred during a lunation or outside, for working 
CCDs or not.

Figure~\ref{fig:effD1_d} compares the model 
for $1-f(t_{0i})$ (blue curve) with the $1-f(t_{0i})$ 
derived directly from  the Monte-Carlo images (shown as the black points).
There is a clear agreement between the two estimates. The model reproduces 
both the efficiency during the observation periods and the loss of efficiency 
between these, which varies with the duration of the absence of data. 
This is clearly emphasized in the bottom plot of Figure~\ref{fig:effD1_d}
where the temporary failures of CCD~3 enhance the efficiency loss due to 
lack of data. 


\section*{Appendix B: additional efficiency plots}
This appendix gives the SN~Ia photometric selection efficiency as a 
function of the 
generated SN~Ia colour and $X_1$.

\begin{figure}[htbp]
\begin{center}
\epsfig{figure=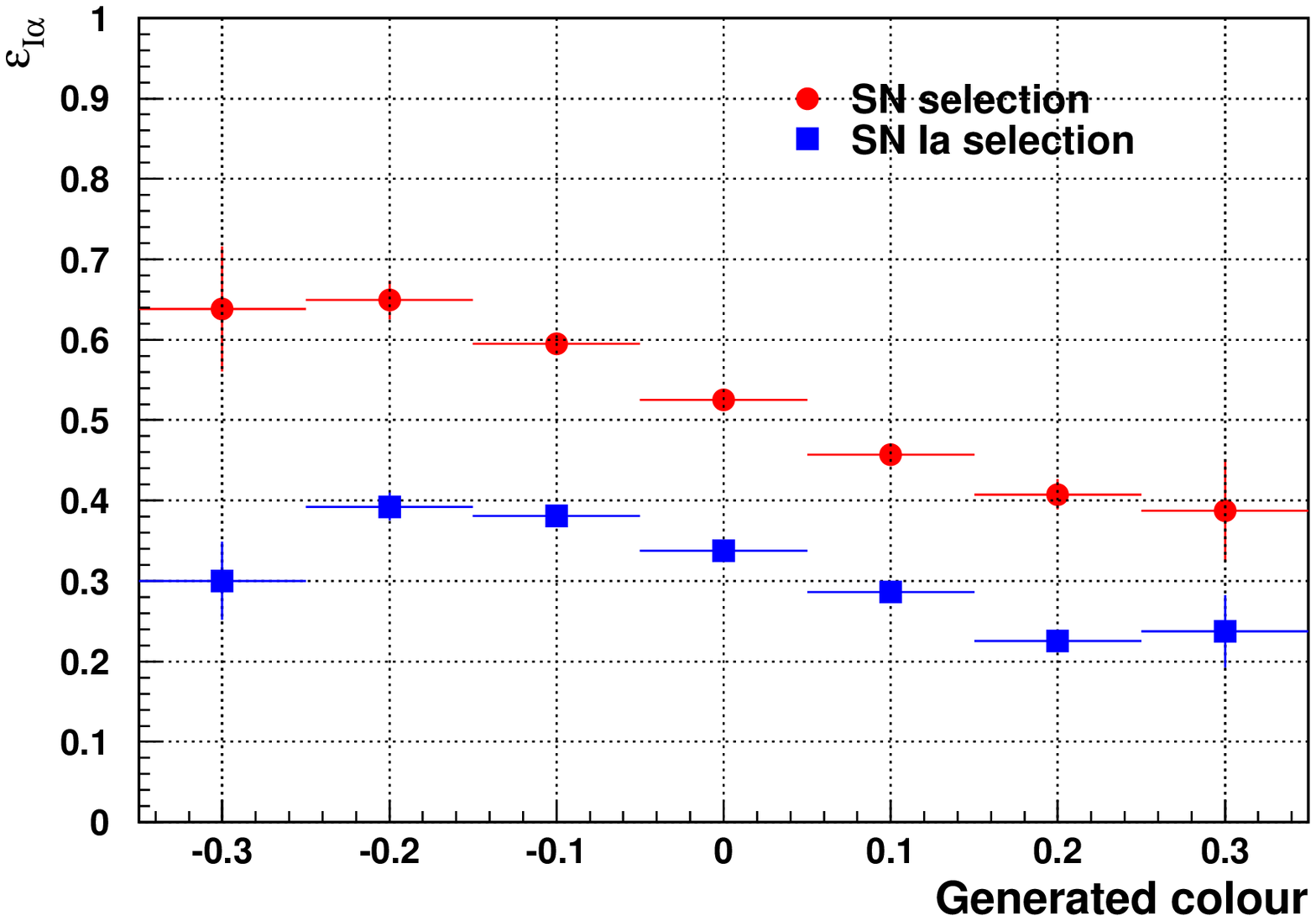, width = \columnwidth} \\
\vspace{-0.5cm}
\epsfig{figure=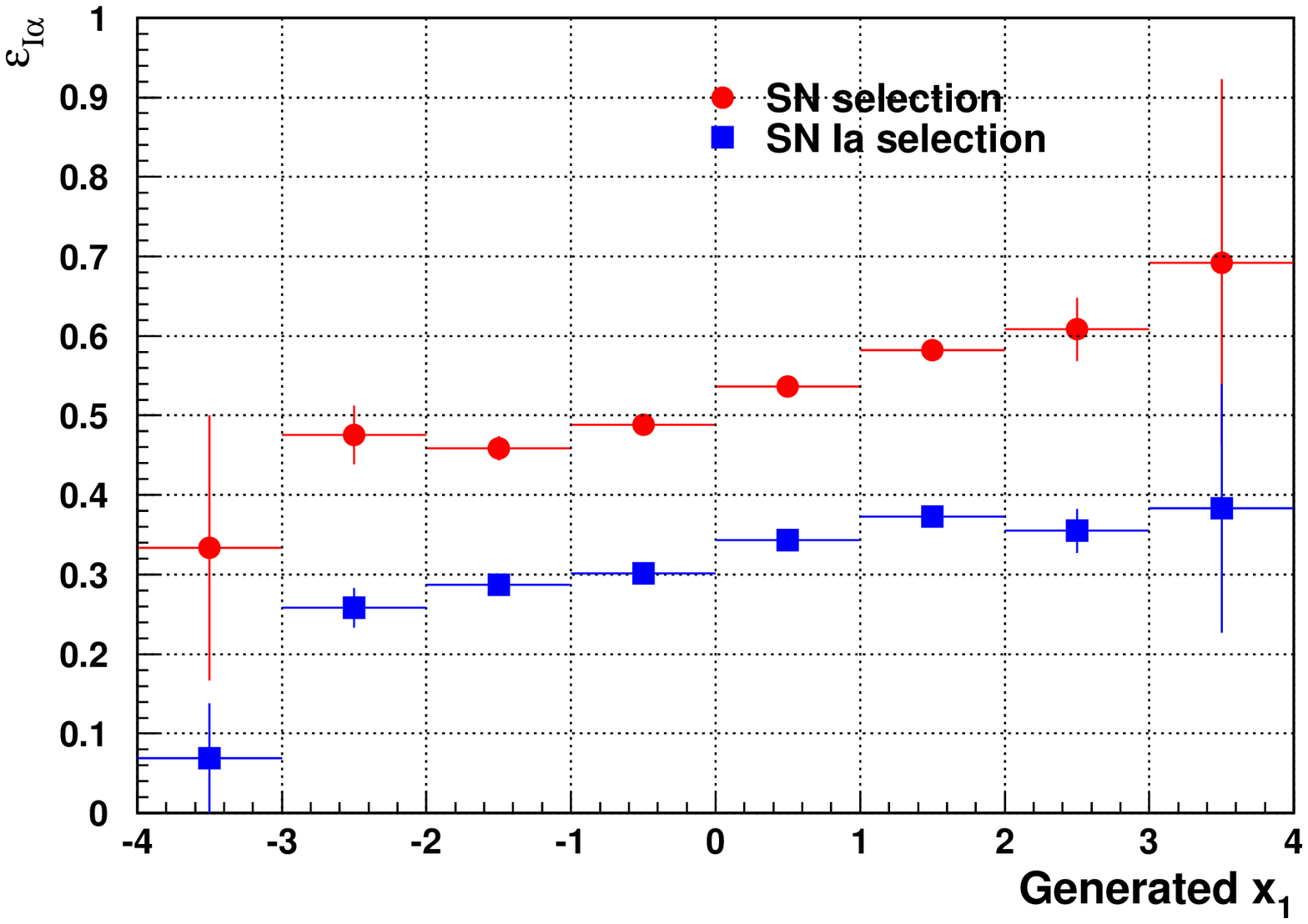, width = \columnwidth} \\
\caption[]{
Selection efficiency from synthetic SN~Ia light curves
as a function of the generated colour and
$X_1$, at different stages of the analysis.} 
\end{center}
\end{figure}

\end{document}